\documentclass[reprint,superscriptaddress,amsmath,amssymb,aps,]{revtex4-2}

\usepackage{graphicx}
\usepackage{dcolumn}
\usepackage{bm}
\usepackage{appendix}
\usepackage{lineno}
\usepackage{threeparttable}
\usepackage{float}
\usepackage{hyperref}
\hypersetup{
  colorlinks=true,    % 着色超链接文本而不是框框
  linkcolor=blue,     % 定义内部链接的颜色
  citecolor=blue,      % 定义引用链接的颜色
  urlcolor=blue       % 定义超链接的颜色
}

\begin{document}

\preprint{APS/123-QED}

\title{Optomechanical preparation of photon number-squeezed states with a pair of thermal reservoirs of opposite temperatures}

\author{Baiqiang Zhu}
\affiliation{State Key Laboratory of Precision Spectroscopy, Department of Physics, School of Physics and Electronic Science, East China Normal University, Shanghai 200062, China}
\affiliation{Shanghai Branch, Hefei National Laboratory, Shanghai 201315, China}

\author{Keye Zhang}
\email{kyzhang@phy.ecnu.edu.cn}
\affiliation{State Key Laboratory of Precision Spectroscopy, Department of Physics, School of Physics and Electronic Science, East China Normal University, Shanghai 200062, China}
\affiliation{Shanghai Branch, Hefei National Laboratory, Shanghai 201315, China}

\author{Weiping Zhang}
\affiliation{Shanghai Branch, Hefei National Laboratory, Shanghai 201315, China}
\affiliation{School of Physics and Astronomy, and Tsung-Dao Lee Institute, Shanghai Jiao Tong University, Shanghai 200240, China}
\affiliation{Shanghai Research Center for Quantum Sciences, Shanghai 201315, China}
\affiliation{Collaborative Innovation Center of Extreme Optics, Shanxi University, Taiyuan, Shanxi 030006, China}

\date{\today}

\begin{abstract}
Photon number-squeezed states are of significant value in fundamental quantum research and have a wide range of applications in quantum metrology. Most of their preparation mechanisms require precise control of quantum dynamics and are less tolerant to dissipation. We propose a mechanism that is not subject to these restraints. In contrast to common approaches, we exploit the self-balancing between two types of dissipation induced by positive- and negative-temperature reservoirs to generate steady states with sub-Poissonian statistical distributions of photon numbers. We also show how to implement this mechanism with cavity optomechanical systems. The quality of the prepared photon number-squeezed state is estimated by our theoretical model combined with realistic parameters for various typical optomechanical systems. 
\end{abstract}

\maketitle

%\tableofcontents
\section{Introduction}

The light field with photon number fluctuations below the standard quantum limit, i.e., the photon number-squeezed state, plays an indispensable role in fundamental research of quantum optics, high-precision metrology, quantum information processing, and other quantum applications~\cite{you2021scalable,thekkadath2020quantum,eaton2023resolution,ansari2018tailoring,madsen2022quantum,goldberg2020transcoherent,goldberg2023beyond}.
The most direct method for preparing number-squeezed states is to give coherent light an intensity-dependent phase shift using a Kerr medium~\cite{PhysRevA.78.023412, PhysRevLett.112.043602, PhysRevA.61.053817,greif2016site,serwane2011deterministic, PhysRevLett.108.183601, PhysRevA.96.053810, PhysRevLett.106.243601, PhysRevLett.107.063601, PhysRevA.85.053802, PhysRevA.100.033822, PhysRevLett.118.223604, doi:10.1126/sciadv.abj1916,yamamoto1992photon}. Sub-Poissonian photon distributions can then be achieved by the resulting equivalent nonlinear photon interaction. Despite the relative simplicity of the implementation, the average photon number and the squeezing degree obtained with this method are limited by the medium.
An alternative approach to generating number-squeezed states involves precise time control or designing complex dynamics processes~\cite{PhysRevLett.125.093603, PhysRevLett.76.1796, PhysRevA.92.040303, PhysRevA.51.1578, Liu_2004, PhysRevLett.115.137002, PhysRevLett.88.143601, PhysRevLett.82.3795,varcoe2000preparing,sayrin2011real, PhysRevA.67.043818, PhysRevA.87.042320, PhysRevLett.127.033602, PhysRevLett.87.093601,hofheinz2008generation,premaratne2017microwave, PhysRevA.80.013805, PhysRevLett.71.3095, PhysRevLett.108.243602,hofheinz2009synthesizing, PhysRevLett.124.063604}, but its low tolerance for noise and time control errors currently restricts the maximum average particle number of squeezed states it prepared to the order of $n\sim 10$.
In addition, the preparation can also be achieved through post-selective measurements of entangled optical modes~\cite{PhysRevLett.71.1816, PhysRevLett.97.073601, PhysRevLett.56.58,guerlin2007progressive, PhysRevA.100.041802, PhysRevA.54.5410, PhysRevLett.70.762, Waks_2006,harder2016single}. This can produce extremely squeezed states, whereas, increasing the average photon number to the order of $n\sim100$, but the production is conditioned on the stochastic measurement results of some other modes, which limits its efficiency.

These approaches are hard to juggle a large average photon number $n$ with high squeezing degree and efficiency because the operating and measurement errors increase rapidly as $n$ rises. In particular, the quantum system considered is essential to have minimal dissipation, and the thermal noise that is always present at finite temperatures has to be eliminated to a large extent as it causes number fluctuations.
A promising solution to the problem of thermal noise in preparing number-squeezed states is quantum-reservoir engineering (QRE)~\cite{PhysRevLett.77.4728}, which harnesses intentional coupling to the environment as a crucial resource of nonclassical steady-state targeting \cite{diehl2008quantum,kraus2008preparation,muschik2011dissipatively}. QRE is less susceptible to experimental noise and in some cases thrives in a noisy environment. 

In this paper, based on QRE with cavity optomechanical coupling, we propose a scheme to generate steady photon number-squeezed states with the help of a pair of positive and negative-temperature optical thermal reservoirs combined with the method of feedback control~\cite{PhysRevA.62.022108}.
The inverse number statistics of two reservoirs have previously been used to drive the heat engine to work with a remarkable efficiency \cite{PhysRevLett.122.240602}, to build a measurement system evading quantum backaction noise \cite{PhysRevA.88.043632}, and to study the emergence of coherence in phase transition dynamics \cite{doi:10.1073/pnas.1408861112}.
Here this characteristic is used to structure sub-Poissionian photon statistics. The mean photon number and the number fluctuation can be changed by the feedback control implemented through dispersive and dissipative optomechanical couplings. 
This leads to considerable simplicity over existing methods and negates any issues involving initial-state preparation, timing control error and coherence time. 
The squeezing quality of the prepared state is decided by the feedback parameters of the specific optomechanical systems but is insensitive to the increase of photon numbers. 
Furthermore, since the squeezed photon number statistics are achieved by the self-feedback balance between two thermal reservoirs, this scheme does not require additional coherent driving and modulation of frequency as in the existing schemes based on QRE \cite{PhysRevLett.115.180501, PhysRevA.93.060301}.
We show that with this scheme, steady states with large mean photon numbers and large degrees of number squeezing are attainable in a variety of cavity optomechanical systems. Additionally, an extremely localized number probability distribution that can be approximated as a number state, is also possible for some systems with very light optomechanical oscillators and strong optomechanical couplings.

\section{method}
The principle of our scheme can be modeled as a harmonic oscillator with nonlinear damping. For instance, for a classical Rayleigh-Van der Por oscillator whose dimensionless dynamical equation is $\ddot{x}+\mu(\dot{x}^2+x^2-1)\dot{x}+x=0$, its damping value depends on the total energy $\dot{x}^2+x^2$. When the energy exceeds $1$, the damping is positive, corresponding to a decay induced by a positive-temperature environment. 
Conversely, when the energy falls below $1$, the damping becomes negative, corresponding to a gain induced by a negative-temperature environment. With this negative feedback, its dynamics finally settle in a limit cycle that satisfies the equation $\dot{x}^2+x^2=1$. 

Similar quantum dynamics can be described by a quantum master equation with energy-dependent dissipation rates. For simplicity, we consider a single-mode bosonic quantum field coupled with a pair of thermal reservoirs of opposite near-zero temperatures, i.e., $T_+\sim 0^+$ and $T_-\sim 0^-$. The Lindblad master equation writes  
\begin{equation}
		\dot{\rho}=-\frac{i}{\hbar}[\hat H_{\rm a},\rho]+\mathcal{D}[\hat{a}\sqrt{\kappa_{\hat{n}}^+}]\rho+\mathcal{D}[\sqrt{\kappa_{\hat{n}}^-}\hat{a}^\dagger]\rho \, ,
		\label{eq:quantum master equation}
\end{equation}
where $\hat H_{\rm a}=\hbar\omega_{\rm a}\hat a^\dagger\hat a$ is the Hamiltonian with the bosonic annihilation operator $\hat a$, and the Lindblad superoperator $\mathcal{D}[\hat{O}]\rho=\hat{O}\rho\hat{O}^\dagger-\frac{1}{2}(\hat{O}^\dagger\hat{O}\rho+\rho\hat{O}^\dagger\hat{O})$ 
with collapse operator $\hat O$ describing the dissipative dynamics induced by thermal reservoirs.  
The expression of $\hat H_a$ reveals the direct relationship between the energy and the quantum excitation number operator $\hat n=\hat a^\dagger\hat a$. So the energy-dependent feedback effect is represented by nonlinear collapse operators $\hat O=\hat{a}\sqrt{\kappa_{\hat{n}}^+}$ for the positive-temperature reservoir and $\hat O=\sqrt{\kappa_{\hat{n}}^-}\hat{a}^\dagger$ for the negative-temperature reservoir, respectively. In particular, the present dissipation rates $\kappa^{\pm}_{\hat n}$ do not depend on the mean excitation number $\bar n$, but on the excitation number operator $\hat n$.

The case of dependence on $\bar{n}$ corresponds to classical feedback control, resulting in the nonlinear evolution equation $\dot{\bar{n}}=\kappa^-_{\bar n}(\bar{n}+1)-\kappa^+_{\bar n}\bar{n}$. Then the steady value of $\bar n$ is controllable, depending on the specific expressions of $\kappa^{\pm}_{\bar n}$, but the steady state is always a thermal equilibrium state. As derived in Appendix A, the steady number fluctuation $\Delta n =\sqrt{\bar{n}^2+\bar{n}}$, which is independent on $\kappa^{\pm}_{\bar n}$. 

By contrast, the case of dependence on $\hat{n}$ corresponds to quantum feedback control which can lead to non-equilibrium steady states, whose number statistics are also controllable.
If the dissipation rates of two reservoirs have opposite variations with excitation number $n$, the steady-state number statistics distribution is determined by two competing dissipative effects.
The probability $P_n$ exponentially decreases versus $n$ in the region dominated by the positive-temperature dissipation and instead exponentially increases in the region dominated by the negative-temperature one. As a result, a peak occurs in the intermediate region, indicating the sub-Poissonian number statistics.

\begin{figure}[t]
	\centering
	\includegraphics[width=0.4\textwidth]{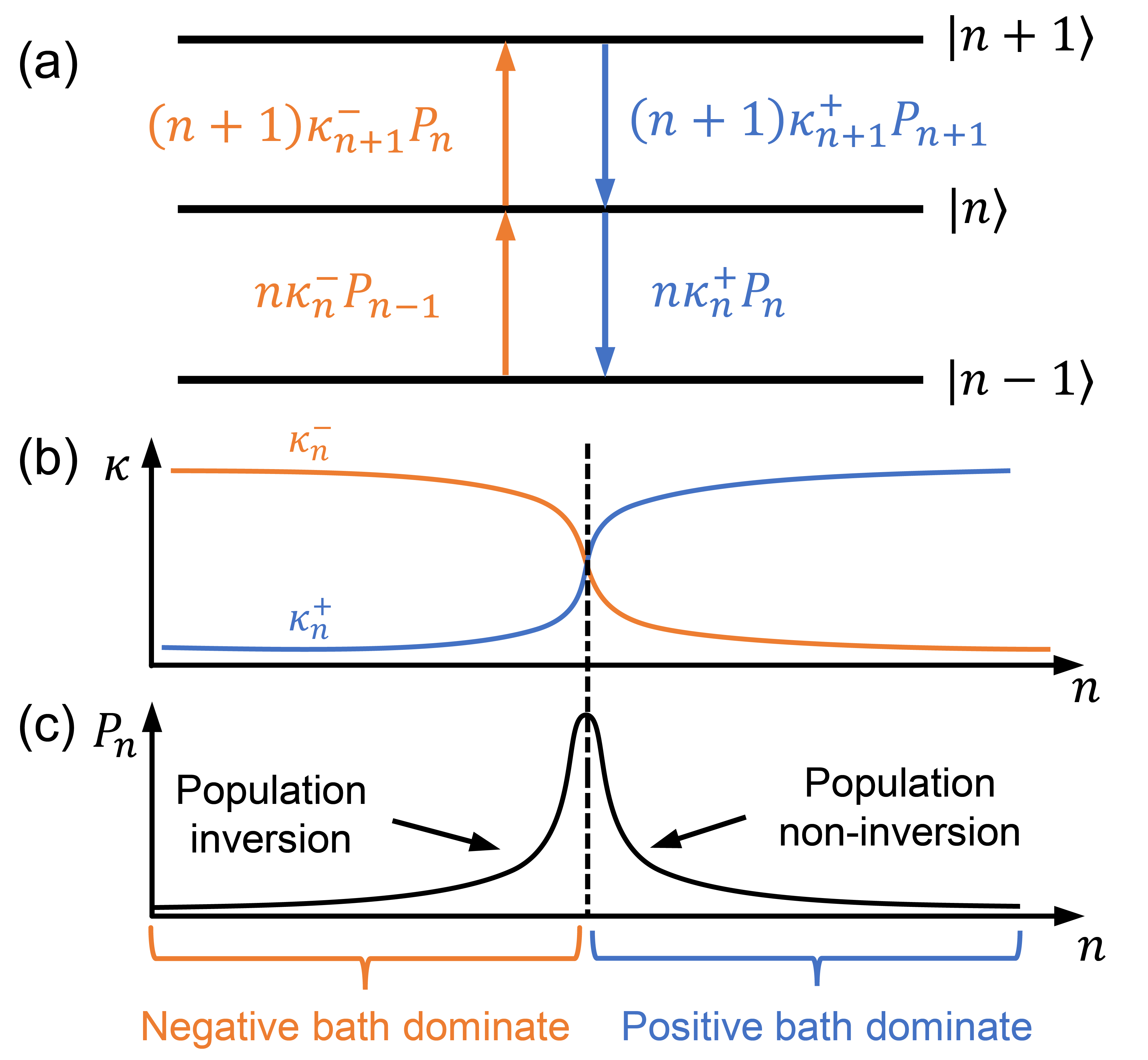}
	\caption{(a) Diagram of the population jump rates between neighboring Fock states. (b) Eigenvalues of the dissipation rate operators $\kappa_{\hat n}^\pm$ versus number $n$. (c) Number statistics distribution of the steady state. The probability $P_n$ increases versus $n$ in the region dominated by the negative-temperature dissipation and decays in the rest region dominated by the positive-temperature one, so a peak appears in the intermediate region.}
	\label{fig:population flow}
\end{figure}

Although nonlinear, the present expressions of collapse operators $\hat{a}\sqrt{\kappa_{\hat{n}}^+}$ and $\sqrt{\kappa_{\hat{n}}^-}\hat{a}^\dagger$ imply that the coupling with the reservoirs is a single quantum interaction, different from the one in multiquantum form as previously used to study quantum nonlinear oscillator~\cite{PhysRevResearch.3.013130}. Then the number statistics of the steady state are readily available by analyzing only the time evolution of the probabilities of neighboring number states. With $\kappa_n^\pm$ representing the eigenvalues of two dissipation rate operators on the number state $|n\rangle$, the evolution equation obtained from the master equation [Eq.~(\ref{eq:quantum master equation})] writes (see Appendix B for details of the derivation),
\begin{equation}
		\dot{P}_n=\kappa_{n+1}^+(n+1)P_{n+1}-\kappa_{n}^+nP_n+\kappa_{n}^-nP_{n-1}-\kappa_{n+1}^-(n+1)P_n \, ,
		\label{eq:diagonal element}
\end{equation}
which, as sketched in Fig.~\ref{fig:population flow}(a), implies that the coupling with positive-temperature reservoir causes a downward jump from state $|n\rangle$ to state $|n-1\rangle$ at a rate $\kappa_{n}^+ n P_n$. 
By comparison, the coupling with negative-temperature reservoir causes an upward jump from state $|n-1\rangle$ to state $|n\rangle$ at a rate of $\kappa_{n}^- n P_{n-1}$. 
The system achieves steady states when the two jumping rates are equal, so the steady number probability distribution is decided by the equation, 
\begin{equation}
	 \frac{P_n}{P_{n-1}}=\frac{\kappa_{n}^-}{\kappa_{n}^+},\,\, n=1, 2, ... +\infty \,.
	 \label{steadyPn}
\end{equation} 

When the dissipation rate $\kappa_n^+$ increases with the number $n$, and instead $\kappa_n^-$ decreases as shown in Fig.~\ref{fig:population flow}(b), a peak in the number probability distribution occurs near the $n$ value satisfying $\kappa_n^+ \approx\kappa_n^-$. For a well-localized single-peak number distribution, the mean value $\bar n$ is very close to the $n$ value. By replacing the discrete distribution with an approximate continuous distribution, see Appendix C, the number fluctuation can be estimated by
\begin{equation}
		\Delta{n}^2\approx \left[ \frac{d}{dn} \left. \left(\frac{\kappa_{n}^+}{\kappa_{n}^-}\right)\right |_{n=\bar n} \right]^{-1}\,,
		\label{eq:variance estimation}
\end{equation}
which implies that to achieve an extremely number-squeezed steady state requires a large derivative of the ratio between two dissipation rates near the peak $n$, which is realizable, for example, when $\kappa_n^+$ and $\kappa_n^-$ have opposite sharp variations there.

The above derivation and conclusion do not rely on the specific expressions of $\kappa_{\hat n}^\pm$. In the following, as a concrete example, we derive the steady number statistics when their expressions are a pair of symmetry logistic functions,
\begin{equation}
		\kappa_{\hat{n}}^\pm=\frac{\kappa_0}{1+\exp\left[\pm{k}(n_0-\hat{n}+\frac{1}{2})\right]} \,,
		\label{eq:control protocol}
\end{equation}
which monotonically change in the interval $[0, \kappa_0]$ with steepness $k$ and midpoint $n_0+1/2$. We will show later these expressions are realizable in optomechanical systems. 

After substituting them into Eq.~(\ref{steadyPn}) and 
iterating, one can obtain the number probability of the steady state as
\begin{equation}
		P_n=e^{k\sum_{i=1}^n(n_0-i+\frac{1}{2})}P_0
		=\mathcal{N}e^{-\frac{k}{2}(n-n_0)^2}\,,
		\label{eq:probability}
\end{equation}
where $\mathcal{N}$ is the normalization factor. 
This indicates a discrete Gaussian-liked probability distribution whose mean number $\bar n \approx n_0$ and variance $\Delta{n}^2\approx 1/k$, respectively, coinciding with the result given by the estimation with Eq.~(\ref{eq:variance estimation}). When the steepness $k>n_0^{-1}$ the steady state is a number-squeezed mixed state with sub-Poissonian number statistics. When $k>1$, this state can be safely approximated as number state $|n_0\rangle$.

\section{Optomechanical implementation}

\begin{figure}[h]
	\centering
	\includegraphics[width=0.46\textwidth]{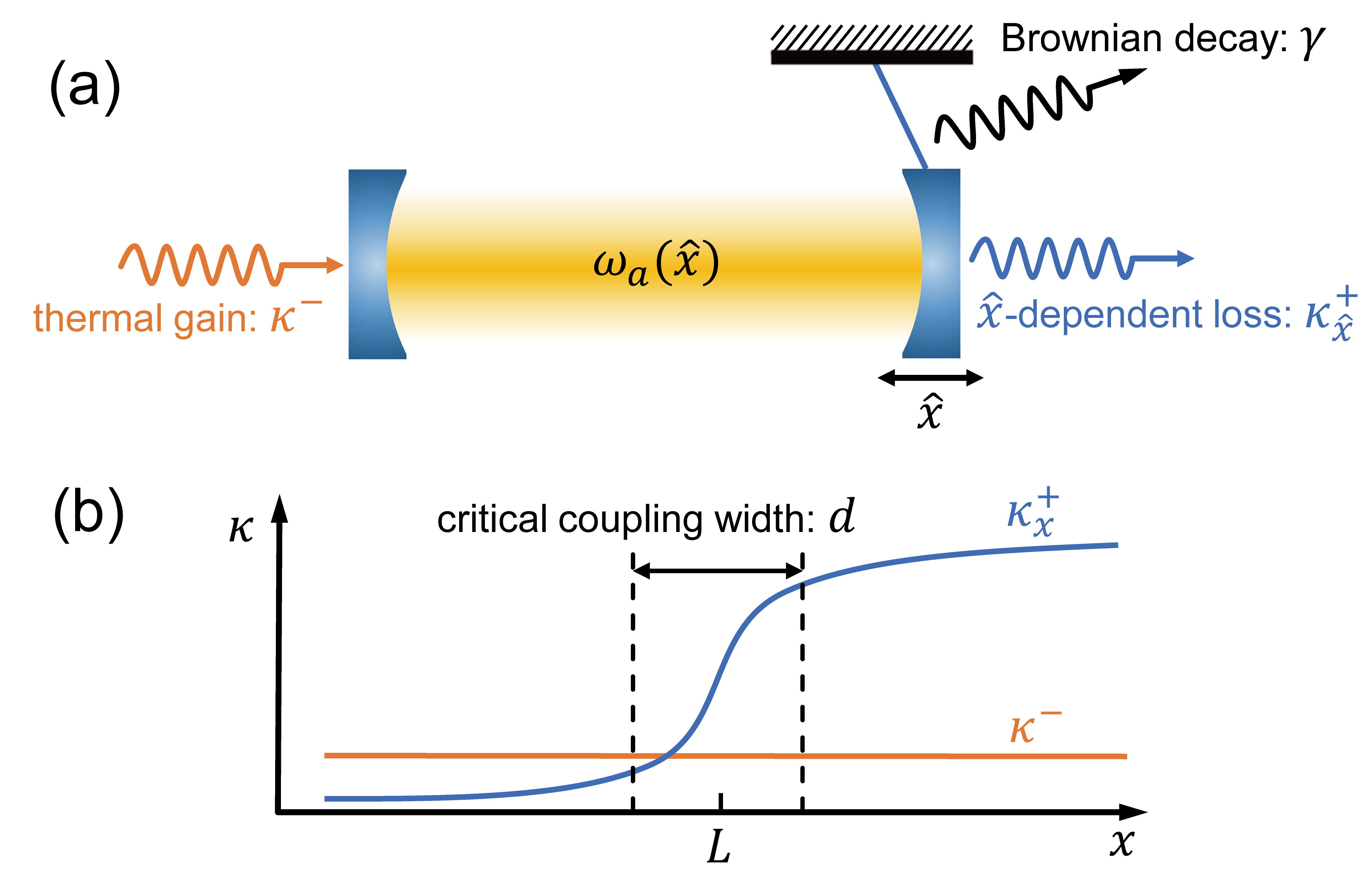}
	\caption{(a) Cavity optomechanical scheme of feedback control. The optical cavity is coupled to the mechanical oscillator through dispersive and dissipative optomechanical interactions simultaneously. With the dispersive coupling, the oscillator undergoes a shift proportional to the radiation pressure force, i.e. to the photon number, and then changes the cavity dissipation rate $\kappa_{\hat{x}}^+$ through the dissipative coupling.
	Except for the optomechanical dissipation, the cavity mode has a gain of rate $\kappa^-$ induced by the negative-temperature reservoir, and the oscillator is subjected to Brownian thermal noise. The high frequency of the optical mode makes our near-zero temperature assumption reasonable.
	(b) Dissipation control protocol: The positive-temperature dissipation rate $\kappa_{\hat{x}}^+$ is smaller than the negative-temperature one in the region $x<L$, but increases rapidly and overtakes it in the region $x>L$. The steep change occurs mainly in a region of width $d$.}
	\label{fig:exp setting}
\end{figure}

Although the above method can be applied to a variety of quantum systems, recent great progress in the research of optomechanical feedback cooling~\cite{kleckner2006sub,li2013millikelvin,magrini2021real,tebbenjohanns2021quantum}, as well as the wide range of feedback parameter due to the diversity of optomechanical structure, suggest that the optomechanical system is more advantageous in implementing this approach.
In what follows, we show an optomechanical scheme for generating number-squeezed states with large photon numbers. As depicted in Fig.~\ref{fig:exp setting}(a), an optical cavity is coupled with a mechanical oscillator through dispersive as well as dissipative optomechanical interactions. 
The optomechanical interaction is dispersive in the sense that the cavity resonance frequency experiences a shift depending on the displacement of the mechanical oscillator arising from photonic radiation pressure. 
Differently, the dissipative optomechanical interaction arises from the dependence of the cavity dissipation rate on mechanical displacement~\cite{PhysRevLett.102.207209}. 
When both interactions come into play, the mechanical oscillator plays the dual role of sensor and controller, sensing the number of photons through the dispersive coupling, and then adjusting the cavity dissipation rate through the dissipative coupling.

The effect of coupling with a negative temperature thermal reservoir is equivalent to introducing a negative optical dissipation, i.e. a gain. This could be provided by quantum dots or other rare-earth-doped media and through nonlinear processes, such as Raman or parametric amplification, which are widely used in the experimental studies of PT-symmetric physics~\cite{chang2014parity, hodaei2017enhanced, el2018non,peng2014parity}. 
In addition, one can also achieve this exotic reservoir with the help of negative-temperature photonic gases recently realized through nonlinear fiber-optic loops~\cite{marques2023observation}.

The total dynamics of the cavity optomechanical system are described by the master equation \cite{PhysRevLett.90.256801, PhysRevB.79.241403, PhysRevB.74.014303}, 
\begin{equation}
	\begin{aligned}
		\dot{\rho}=-\frac{i}{\hbar}[\hat{H}_{\rm tot},\rho]+\mathcal{D}[\sqrt{\kappa_{\hat{x}}^+}\hat{a}]\rho+\kappa^-\mathcal{D}[\hat{a}^\dagger]\rho+\mathcal{L}_{\rm m}\rho\,,
		\label{eq:exp master equation}
	\end{aligned}
\end{equation}
where the Hamiltonian
\begin{equation}
	\hat{H}_{\rm tot}=\hbar(\omega_{\rm a}-G\hat{x})\hat{a}^\dagger\hat{a}+\frac{\hat{p}^2}{2m}+\frac{1}{2}m\omega_{\rm m}^2\hat{x}^2 \,.
\end{equation}

The present $\hat a$ represents the annihilation operator of the optical cavity mode at frequency $\omega_{\rm a}$, $\hat x$ and $\hat p$ are the position and momentum operators of the mechanical oscillator, with mass $m$ and frequency $\omega_{\rm m}$. The dispersive optomechanical coefficient $G=g_0/x_{\rm zpf}$ with $g_0$ the vacuum optomechanical coupling strength and $x_{\rm zpf}=\sqrt{\hbar/2m\omega_{\rm m}}$ the zero-point fluctuation amplitude. 

The dissipative coupling leads to a displacement-dependent modulation of the cavity dissipation rate. The dependence is commonly assumed to be linear when the displacement is small~\cite{PhysRevLett.107.213604,tagantsev2021dissipative}. When the displacement range is large the dependence becomes nonlinear. In general, the dissipation rate only changes rapidly within a certain displacement range, whereas, it changes slowly when the displacement is too small or too large. The exact variation depends on the property of the specific optomechanical system, see, for example, Fig. 3(a) of Ref.~\cite{sankey2010strong}. Considering the different variations in each part, we fit the $x$ dependence of the dissipation rate with a logistic function,
\begin{equation}
	\kappa_{\hat{x}}^+=\kappa_{\rm v}+\frac{\kappa_0}{1+\exp(4(L-\hat{x})/d)},
	\label{kappaxplus}
\end{equation}
where $\kappa_{\rm v}$ and $\kappa_0$ represent the original vacuum dissipation rate and the amplitude of the modulation, respectively. $L$ is the critical coupling distance indicating the displacement value at which the dissipative rate has the fastest change, and $d$ is the coupling width indicating the displacement range where the dissipation rate changes significantly.

For simplicity, here we consider only a constant negative-temperature dissipation rate $\kappa^-$.
As discussed above a probability peak at $n_0$ occurs in the number statistics when the negative-temperature dissipation dominates on the side $n<n_0$ and the positive-temperature dissipation dominates the other side. Considering the mechanical displacement $\hat x$ is proportional to the photon number $\hat n$, an ideal variation of cavity dissipation rate is, as shown in Fig.~\ref{fig:exp setting}(b), that $\kappa^+_{x}\ll \kappa^-$ in the region $x<L-d/2$ but in the region  $x>L+d/2$, $\kappa^+_{x}$ increases rapidly with $x$ and eventually dominates.

The dissipation of the mechanical oscillator is also considered in the form of Brownian thermal noise, described by the superoperator
\begin{equation}
		\mathcal{L}_{\rm m}\rho=-\frac{i\gamma}{2\hbar}[\hat{x},\{\hat{p},\rho\}]-\frac{\gamma}{2}(n_{\rm th}+\frac{1}{2})[\hat{x},[\hat{x},\rho]]\,,
\end{equation}
where $\gamma$ and $n_{\rm th}$ are the mechanical dissipation rate and the thermal mean phonon number of the oscillator, respectively.

\begin{table*}[htbp]\small
	\centering
	\caption{\bf Experimental parameters and ideal squeezing degrees for several representative optomechanical systems.}
	\renewcommand\arraystretch{1.25}
	\begin{threeparttable}
	\begin{tabular*}{1\linewidth}{llllllllll}
		\hline
		Setup & $m_{\rm eff}({\rm kg})$ & $\omega_{\rm m}/2\pi({\rm Hz})$ & $g_0/2\pi({\rm Hz})$ & $x_1(\rm nm)$ & $d({\rm nm})$ & $L({\rm nm})$ & $\Delta{n}$ & $\bar{n}$ & $\Delta n^2/\bar{n}(\rm dB)$ \\
		\hline
		Micromirror \cite{kleckner2011optomechanical} & $1.1\times10^{-10}$ & $9.7\times10^{3}$ & $22$ & $1.27\times10^{-8}$ & $2.48^*$ & $50$ \cite{favero2007optical} & $7\times10^{3}$ & $4\times10^{9}$ & $-19$ \\
		SiN membrane \cite{PhysRevLett.107.213604} & $1\times10^{-10}$ & $1.03\times10^{5}$ & $0.57$ & $1\times10^{-11}$ & $2.48$ & $100$ \cite{thompson2008strong} & $2.5\times10^{5}$ & $1\times10^{13}$ & $-22$ \\
		Micro-disk \cite{PhysRevLett.103.223901} & $2\times10^{-15}$ & $2.5\times10^{7}$ & $26$ & $2.6\times10^{-11}$ & $0.04$ & $0.02$ \cite{PhysRevLett.94.223902} & $1.9\times10^{4}$ & $7.7\times10^{8}$ & $-3$ \\
		Levitated particle \cite{delic2020cooling} & $2.8\times10^{-18}$ & $3\times10^{5}$ & $3$ \cite{ranfagni2021vectorial,romero2010toward} & $6.3\times10^{-8}$ & $0.3^*$ & $30$ \cite{PhysRevLett.129.013601,PhysRevLett.122.223602} & $1.1\times10^3$ & $4.8\times10^8$ & $-26$ \\
		Photonic crystal \cite{PhysRevX.4.021052} & $4\times10^{-16}$ & $4.9\times10^{6}$ & $1.3\times10^{5}$ & $3.5\times10^{-6}$ & $10$ & $100$ & $8.4\times10^{2}$ & $2.9\times10^{7}$ & $-16$ \\
		%BPT molecule \cite{PhysRevLett.125.233601} & $4.9\times10^{-29}$ & $4.7\times10^{13}$ & $3.3\times10^{10}$ & $8.5\times10^{-5}$ & $100$ & $0.1$ & $5.4\times10^{2}$ & $1.2\times10^{3}$ & $23.98$ \\
		Cold atomic gases \cite{PhysRevLett.105.133602} & $2.4\times10^{-22}$ & $7\times10^{4}$ & $3.5\times10^{6}$ & $70$ & $25^*$ & $2500$ & $0.3$ & $34.8$ & $-26$ \\
		\hline
	\end{tabular*}
	\end{threeparttable}
	\label{tab:exp parameters}
\end{table*}

Below we display that a number-state-sensitive optical dissipation rate and then a steady photon number-squeezed state as we proposed above are obtained in the limit of large mechanical dissipation, i.e. $\gamma\gg\kappa^\pm$. 
In this limit the oscillator adiabatically follows the slowly varying optical field, acting as a quick-response feedback control unit. The expression of the optomechanical steady state and the photon number statistics can be derived analytically from the master equation [Eq.~(\ref{eq:exp master equation})] with the adiabatic approximation. We present the main results here and place the detailed derivations in Appendix D.

Considering the optomechanical oscillator is trapped by a displaced harmonic potential, $m\omega_{\rm m}^2\hat{x}^2/2-\hbar G\hat{n}\hat{x}$, the state of the oscillator $\rho_{\rm m}(n)$ closely approximates a thermal state with a photon number $n$-dependent displacement. 
This state then leads to an $n$-dependent cavity dissipation rate of the form similar to Eq.~(\ref{eq:control protocol}),
\begin{equation}
	\kappa_{n}^+={\rm Tr}[\kappa_{\hat{x}}^+\rho_{\rm m}(n)]\approx\kappa_{\rm v}+\frac{\kappa_0}{1+\exp\left[4(L-n x_1)/d^{\prime}\right]}\,,
	\label{eq: exp control protocol}
\end{equation} 
where $x_1=2g_0x_{\rm zpf}/\omega_{\rm m}$, representing the displacement of the oscillator under the radiation pressure force generated by a single photon, and $d'=\Delta{x}/\tanh(\Delta{x}/d)$, indicating that the effective coupling width $d$ is blurred by the thermal fluctuation of position whose expression is $\Delta{x}=x_{\rm zpf}\sqrt{2n_{\rm th}+1}$.

The other dissipation rate $\kappa^-$ for the negative-temperature reservoir is constant, but the photon number statistics of the steady state can still present a single-peak distribution as long as the dissipation ratio $\kappa^-/\kappa_n^+$ changes rapidly with $n$.
The mean photon number and the fluctuation can be estimated by Eqs.~(\ref{eq:control protocol}) and (\ref{eq:probability}), which give
\begin{eqnarray}
		\bar n&\approx&\frac{L}{x_1}-\frac{1}{2}\, ,\label{eq: mean}\\
		\Delta{n}&\approx&\sqrt{\frac{d'}{4x_1}}\,.
		\label{eq: fluctuation}
\end{eqnarray}

So to achieve a photon number-squeezed distribution with a large $\bar n$ but a small $\Delta n$, a large critical coupling distance $L$ and a small coupling width $d$ are required at the same time, which means a sharp variation of dissipation rate $\kappa_x^+$ takes place after a large photon-pushed displacement. A smaller positional fluctuation $\Delta x$ of the oscillator is always better because it is more favorable to obtain a smaller photon number fluctuation through optomechanical feedback. However, the case is different for the single-photon displacement $x_1$, because its decrease results in the increases of $\bar n$ and $\Delta n$ simultaneously. 

The specific performance of the number-squeezed-state preparation depends on the parameters of the optomechanical system being used.
For reference, in Table \ref{tab:exp parameters} we evaluate the steady photon number statistics and the squeezing degree of this scheme in the ideal adiabatic case with the practical parameters of several different optomechanical systems.
The critical coupling distance $L$ is determined by the optomechanical displacement value at which the change rate $\partial\kappa/\partial x$ reaches its maximum. The coupling width $d$ is determined by the difference between two special displacement values at which the second derivative $\partial^2\kappa/\partial x^2$ reaches its maximum and minimum, respectively. The starred data are estimated values.
Most of these systems can achieve a very high photon number squeezing degree, defined by $\Delta n^2/\bar{n}$, but the mean photon number $\bar{n}$ and the number fluctuation $\Delta n$ vary significantly between systems. For some systems with small oscillator mass $m_{\rm eff}$ but strong optomechanical coupling $g_0$, such as cold atoms, the steady photon number fluctuation can become very small due to their large single-photon displacements $x_1$, and the resulting squeezed state is close to a large-$n$ number state.

However, in realistic experiments, there are several challenges in achieving these impressive squeezing degrees. The first is the adiabatic limit in which the oscillator reacts quickly to the changes in the radiation pressure force. This could be reached by increasing the mechanical dissipation rate $\gamma$, but the accompanied large position fluctuations will prevent the oscillator from having a photon-number-resolved response unless the single-photon coupling strength $g_0$ is large enough. 
The second is the requirement for dramatic variations in the cavity dissipation rate. Such variation can be achieved in systems with large dissipative optomechanical coupling, but large rates and ranges in variation are often difficult to achieve simultaneously. One can only make trade-offs based on the specific system.

\begin{figure}[ht]
	\centering
	\includegraphics[width=0.4\textwidth]{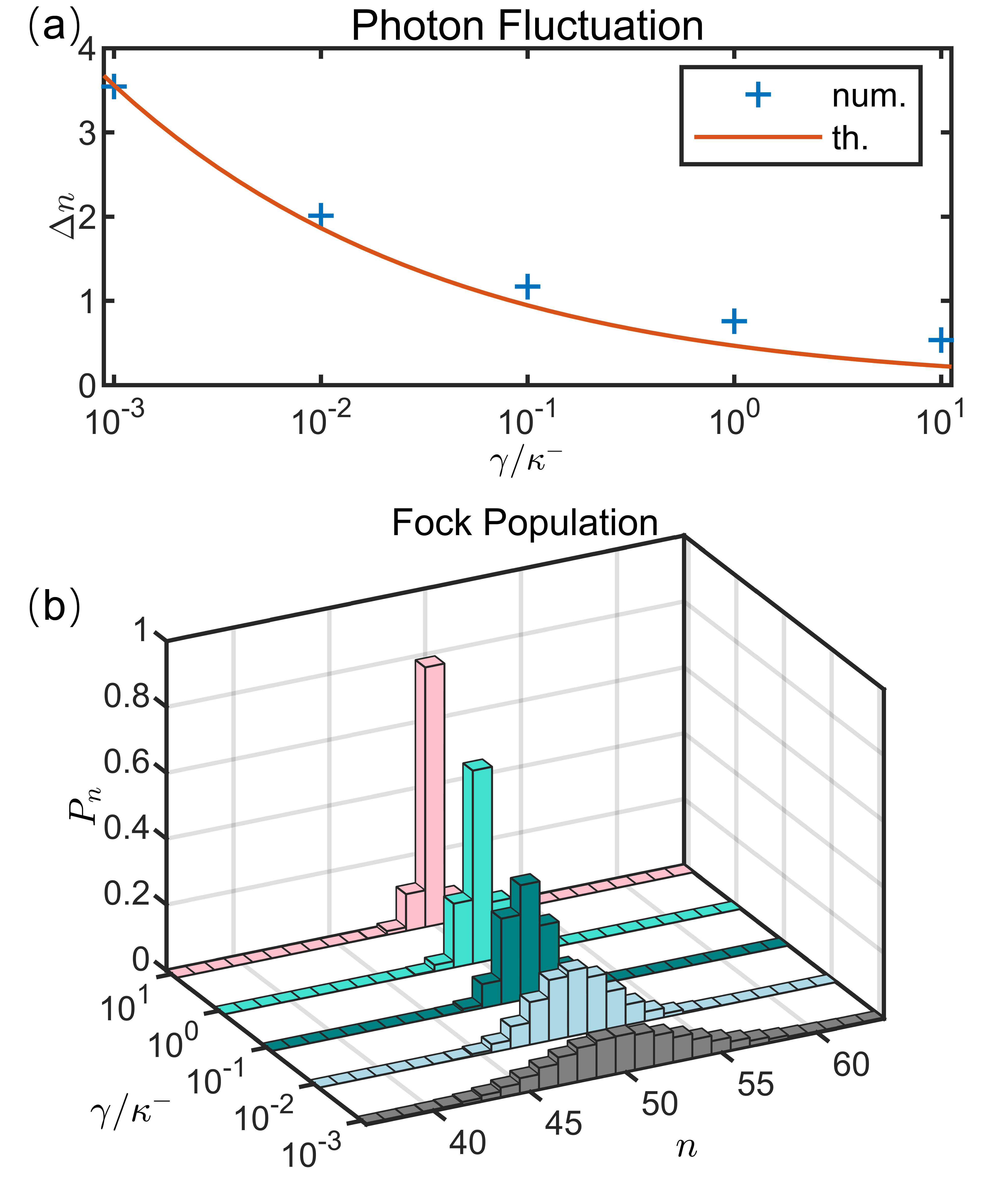}
	\caption{Steady-state photon number statistics obtained by approximate solution [Eq.~(\ref{eq: fluctuation})] and numerical simulation of the master equation [Eq.~(\ref{eq:exp master equation})]. (a) Photon number fluctuation $\Delta n$ versus dissipation ratio $\gamma/\kappa^-$. The approximate solution is plotted in a red solid line while the numerical results are marked with a "$+$".
		(b) Numerical results for the steady-state probability distribution of photon number for increasing $\gamma/\kappa^-$. 
		All results for $g_0=7.07\times10^{2}\omega_{\rm m}$, $(\kappa_0, \kappa^-, \kappa_{\rm v})=(10^{-1},10^{-2},10^{-3})\omega_{\rm m}$, and $(d, L)=(14, 7\times10^4)x_{\rm zpf}$.}
	\label{fig:exp transition region}
\end{figure}

Finally, as an illustration, we analyzed the steady photon number statistics when the adiabatic condition $\gamma\gg\kappa^\pm$ is not well satisfied. In this case, the feedback response is slow. The backaction of the feedback control, carried out by dissipative optomechanical interaction, prevents the state of the mechanical oscillator from being approximated as a mixture of several $n$-dependent displaced thermal states.    
This (see Appendix D for a detailed derivation) leads to an increase in the effective optomechanical coupling width followed by a modification of the cavity dissipation rate in Eq.~(\ref{eq: exp control protocol}), 
\begin{equation}
	\kappa_{n}^+ \approx\kappa_{\rm v}+\frac{\kappa_0}{1+\exp\left[4\xi(L-nx_1)/d'\right]}\,,
	\label{eq:exp control protocol1}
\end{equation}
which next leads to a modification of the steady-state photon number fluctuation in Eq.~(\ref{eq: fluctuation}),
\begin{equation}
	\Delta{n}\approx\sqrt{\frac{d'}{4\xi x_1}}\,,
	\label{eq: fluctuation1}
\end{equation}
where the modification factor,	
\begin{equation}
\xi=1-\frac{1}{\cosh(\sqrt{\frac{\Delta n^2\gamma}{\bar{n}\kappa^-}})+\sqrt{\frac{\gamma}{\bar{n}\kappa^-}}\sinh(\sqrt{\frac{\Delta n^2\gamma}{\bar{n}\kappa^-}})}.
	\label{eq:slope variation}
\end{equation}

The approximate value of steady-state photon number fluctuation $\Delta n$ can be obtained by solving Eqs.~(\ref{eq: fluctuation1}) and (\ref{eq:slope variation}). In Fig.~\ref{fig:exp transition region}(a) we show the solution as a function of the ratio $\gamma/\kappa^-$. The fluctuation increases as the ratio decreases. In the limit case $\gamma\ll\kappa^-$ we have $\xi\approx 0$, and the response of the oscillator is too slow to act as a feedback unit. The cavity dissipation rate tends to be a constant, $\kappa_{n}^+\approx\kappa_v+\kappa_0/2$, and then the steady-state photon number distribution tends to be thermal.
For comparison, in Fig.~\ref{fig:exp transition region}(a) we also label several exact values of $\Delta n$ obtained by numerically solving the full master equation [Eq.~(\ref{eq:exp master equation})]. They fit well with the approximate solutions. The exact photon number probabilities $P_n$ are shown in Fig.~\ref{fig:exp transition region}(b). For an identical mean photon number, the localization of the number probability distribution becomes more and more significant as the ratio $\gamma/\kappa^-$ increases, eventually converging to a definite number, i.e., a photon number state.

\section{conclusions}
To summarize, we proposed a method to deterministically generate photon number-squeezed states based on feedback control and reservoir engineering techniques. The method did not require precise timing control of the quantum dynamics and was tolerant to noisy environments. As a demonstration, we proposed an implementation scheme with cavity optomechanical systems. The significant photon number squeezing of the steady state stemmed from the cooperation between a pair of positive- and negative-temperature optical thermal reservoirs with a feedback controller played by an optomechanical oscillator. Thanks to the diversity of the optomechanical system, its mean photon number and number fluctuation can be tuned in a wide range, even up to approximate number states with high photon numbers. It would be interesting to inquire whether the other special number probability distributions can be realized by this approach. For example, a nonmonotonic variation of dissipation rate with the optomechanical displacement could lead to a multipeaked probability distribution of the photon number. Other future directions include considering more than one optical mode and other implementations besides optomechanical systems.

\section{Acknowledgments}
We acknowledge ﬁnancial support from the Innovation Program for Quantum Science and Technology (Grant No. 2021ZD0303200), the National Key Research and Development Program of China (Grant No. 2016YFA0302001), the National Science Foundation of China (Grants No. 11974116, No. 12234014, and No. 11654005), the Shanghai Municipal Science and Technology Major Project (Grant No. 2019SHZDZX01), and the Fundamental Research Funds for the Central Universities. K. Z. acknowledges the Chinese National Youth Talent Support Program. W. Z. also acknowledges additional support from the Shanghai talent program.

\appendix

\section{Steady state under classical feedback control}
Under the assumption of classical feedback control, the dissipation rate is no longer a quantum operator but a value that depends on the mean excitation number $\bar{n}$, that is, $\kappa_{\hat n}^{\pm}\rightarrow \kappa_{\bar n}^{\pm}$. 
Then according to the master equation [Eq.~(\ref{eq:quantum master equation})], one can obtain the time evolution equations for the first and second-order moments of the excitation number operator,
\begin{eqnarray}
		\frac{d\langle\hat{n}\rangle}{dt}&=&\kappa_{\bar n}^-(\langle\hat{n}\rangle+1)-\kappa_{\bar n}^+\langle\hat{n}\rangle\,,\\
		\frac{d\langle\hat{n}^2\rangle}{dt}&=&\kappa_{\bar n}^-(\langle\hat{n}\rangle-2\langle\hat{n}^2\rangle)+\kappa_{\bar n}^+(2\langle\hat{n}^2\rangle+3\langle\hat{n}\rangle+1)\,,
		\label{eq:first and second moment}
\end{eqnarray}
which, in the steady-state case lead to the equation $\langle\hat{n}^2\rangle=2\langle\hat{n}\rangle^2+\langle\hat{n}\rangle$. So the normalized second-order correlation function $g^{(2)}=2$, which implies that the steady state is a thermal state.

\section{Dynamical equation for number-state population}
From the master equation [Eq.~(\ref{eq:quantum master equation})], we can derive the evolution equations for the population on each Fock state, 
\begin{equation}
	\dot{P}_n=\langle n|\dot{\rho}|n\rangle = \langle n|(-\frac{i}{\hbar}[\hat H_{\rm a},\rho]+\mathcal{D}[\hat{a}\sqrt{\kappa_{\hat{n}}^+}]\rho+\mathcal{D}[\sqrt{\kappa_{\hat{n}}^-}\hat{a}^\dagger]\rho) |n\rangle\, ,
\end{equation}
where the term $\langle n|[\hat{H}_{\rm a},\rho]|n\rangle=0$, and considering the eigenequations of the dissipative rate operators, i.e., $\kappa_{\hat{n}}^\pm|n\rangle=\kappa_{n}^\pm|n\rangle$, and the formulas $\hat{a}|n\rangle=\sqrt{n}|n-1\rangle$ and $\hat{a}^\dagger|n\rangle=\sqrt{n+1}|n+1\rangle$, the terms of the Lindblad super operators contributes as follows
\begin{eqnarray}
	\langle n|\hat{a}\sqrt{\kappa_{\hat{n}}^+}\rho\sqrt{\kappa_{\hat{n}}^+}\hat{a}^\dagger|n\rangle&=&\kappa_{n+1}^+(n+1)\langle n+1|\rho|n+1\rangle,\\
	\langle n|\sqrt{\kappa_{\hat{n}}^+}\hat{a}^\dagger\hat{a}\sqrt{\kappa_{\hat{n}}^+}\rho|n\rangle&=&\kappa_{n}^+n\langle n|\rho|n\rangle,\\
	\langle n|\sqrt{\kappa_{\hat{n}}^-}\hat{a}^\dagger\rho\hat{a}\sqrt{\kappa_{\hat{n}}^-}|n\rangle&=&\kappa_n^-n\langle n-1|\rho|n-1\rangle,\\
	\langle n|\hat{a}\kappa_{\hat{n}}^-\hat{a}^\dagger\rho|n\rangle&=&\kappa_{n+1}^-(n+1)\langle n|\rho|n\rangle.
\end{eqnarray}
Then we obtain the time evolution of $P_n$ in the form shown in Eq.~(\ref{eq:diagonal element}).

\section{Estimation of number fluctuation}
According to the parameter estimation theory, the uncertainty of the parameter to be estimated
is equivalent to the peak width of the likelihood function $\mathcal{L}(\theta|x)$.
For a single-peaked probability distribution $P(x)$, if one shifts the random variable $x$ to $x+\theta$, then the peak width of the likelihood function of the parameter $\theta$ is equal to the peak width of the probability distribution of the random variable $x$, i.e. 
\begin{eqnarray}
	{\rm var}(x)&=&{\rm var}(\theta)\nonumber\\
	&=& -[\partial^2_\theta \ln \mathcal{L}(\theta|x_i)|_{\theta=x_0-x_i}]^{-1}\nonumber\\
	&=&-[\partial^2_\theta \ln P(x_i+\theta)|_{\theta=x_0-x_i}]^{-1}\,,
\end{eqnarray}
where $x_i$ is a member of the sampling and $x_0$ is the peak value point of the distribution $P(x)$. With the substitution $x=\theta+x_i$ we have ${\rm var}(x)=-[\partial^2_x \ln P(x)|_{x=x_0}]^{-1}$.
So the variance of the single-peaked probability distribution $P(n)$ can be estimated by $1/\Delta n^2=-\partial^2_n \ln P(n)|_{n=\bar{n}}$.

It should be noted that this derivation is only applicable in a single-peaked probability distribution because the likelihood function corresponding to a single sampling is also single-peaked, and the peak width is equal to the variance of the random variable.

\section{Steady state of the optomechanical feedback control system}
%\subsection{Fast response case}
Considering there is no coherent input to the cavity mode but only decoherence induced by thermal dissipation, the steady state of the cavity optomechanical system can be expressed as
\begin{equation}
	\rho_{\rm s}=\sum_nP_n|n\rangle\langle n |\otimes\rho_{\rm m}(n)\,,
\end{equation}
where $|n\rangle$ represents the $n$-photon number state with probability $P_n$ and $\rho_{\rm m}(n)$ is the density matrix of the mechanical oscillator that depends on photon number $n$. 

To determine the values of $P_n$, we substitute a state of this form into the master equation [Eq.~(\ref{eq:exp master equation})]. After tracing out the part of the mechanical oscillator, we obtain
\begin{equation}
		\dot{P}_n=(n+1)P_{n+1}\kappa_{n+1}^+ -nP_n\kappa_{n}^+
		+nP_{n-1}\kappa^--(n+1)P_{n}\kappa^-\, ,
		\label{eq:reduced master equation1}
\end{equation}
where $\kappa_{n}^+={\rm Tr}[\kappa_{\hat{x}}^+\rho_{\rm m}(n)]$, representing the mean value of the dissipation rate operator under the oscillator state $\rho_{\rm m}(n)$. Similarly, the evolution equation of $\rho_{\rm m}(n)$ can be obtained from the master equation [Eq.~(\ref{eq:exp master equation})] by tracing out the part of the photon, 
\begin{eqnarray}
	\dot{\rho}_{\rm m}(n)&&=-\frac{i}{\hbar}[\hat{H}_n,\rho_{\rm m}(n)]+\mathcal{L}_{\rm m}\rho_{\rm m}(n)
	\nonumber\\
	&&+\frac{P_{n+1}}{P_n}(n+1)\kappa_{n+1}^+(\rho_{\rm m}(n+1)-\rho_{\rm m}(n))
	\nonumber\\
	&&+\frac{P_{n-1}}{P_n}n\kappa^-(\rho_{\rm m}(n-1)-\rho_{\rm m}(n))\,,
	\label{eq:reduced master equation2}
\end{eqnarray} 
where $\hat{H}_{n}=\hat{p}^2/2m+m\omega_{\rm m}^2\hat{x}^2/2-\hbar Gn\hat{x}$, indicating the effective Hamiltonian of the oscillator driven by the optical force generated by $n$ photons. The second term on the right side of the equation represents the damping induced by mechanical dissipation, whereas, the last two terms are induced by dissipative optomechanical interaction and represent the backaction of the feedback control.

\emph{Fast feedback limit}---When $\gamma\gg\kappa^\pm$, the dissipation term $\mathcal{L}_{\rm m}\rho_{\rm m}(n)$ dominates in Eq.~(\ref{eq:reduced master equation2}), and the backaction of feedback control is negligible. Then the state of the oscillator can be approximated as a thermal state with a photon number $n$-dependent displacement, that is, $\rho_{\rm m}(n)\approx D(g_0n/\omega_{\rm m})\rho_{\rm th}D^\dagger(g_0n/\omega_{\rm m})$ where the displacement operator $D(\alpha)=\exp(\alpha\hat{b}^\dagger-\alpha^*\hat{b})$ with $\hat{b}$ the phonon annihilation operator of the mechanical oscillator. With this approximated expression, the $n$-dependent displacement $x_n={\rm Tr}[\hat{x}\rho_{\rm m}(n)]\approx n x_1$ and the $n$-dependent dissipation rate is decided by the integral, 
\begin{equation}
	\kappa_{n}^+ ={\rm Tr}[\kappa_{\hat{x}}^+\rho_{\rm m}(n)]\approx \int\left(\kappa_{\rm v}+\frac{\kappa_0}{1+\exp(4(L-x)/d)}\right) p_n(x){\rm d}x\,,
	\label{kappan}
\end{equation}
where the $n$-dependent position probability
\begin{equation}
	p_n(x)=\langle{x}|\rho_{\rm m}(n)|{x}\rangle\approx\frac{1}{\sqrt{2\pi}\Delta{x}}\exp\left(-\frac{(x-x_n)^2}{2\Delta{x}^2}\right)\,,
	\label{pn}
\end{equation}
with the thermal position fluctuation defined as $\Delta{x}=x_{\rm zpf}\sqrt{2n_{\rm th}+1}$. If the optomechanical coupling width $d\gg\Delta x$, the integral has an approximate expression,
\begin{equation}
	\kappa_{n}^+ \approx \kappa_{\rm v}+\frac{\kappa_0}{1+\exp(4(L-nx_1)/d')}\,.
\end{equation}
This $n$-dependent dissipation rate leads to a non-equilibrium steady state of the optical mode.

\emph{Slow feedback case}---When $\gamma\sim\kappa^\pm$, the last two terms on the right side of Eq.~(\ref{eq:reduced master equation2}) are no longer negligible. The displacement of the oscillator depends not only on the photon number but also on the number fluctuation. Due to the complexity of Eq.~(\ref{eq:reduced master equation2}), an analytical solution is difficult to obtain, but we can derive the evolution equation for the mean displacement,
\begin{eqnarray}
	\dot{x}_n={\rm Tr}[\hat{x}\dot\rho_{\rm m}(n)]&&=-\gamma(x_n-nx_1)
	\nonumber\\
	&&+\frac{P_{n+1}}{P_n}(n+1)\kappa_{n+1}^+(x_{n+1}-x_n)
	\nonumber\\
	&&+\frac{P_{n-1}}{P_n}n\kappa^-(x_{n-1}-x_n)\,,
	\label{eq:reduced master equation3}
\end{eqnarray}  
where we have adiabatically eliminated the equation of the momentum to focus on the coupling between the photon number and the displacement, and $x_1=2g_0x_{\rm zpf}/\omega_{\rm m}$, representing the single-photon displacement of the oscillator.

An approximate solution to Eq.~(\ref{eq:reduced master equation3}) is given by $x_n=x_1 \bar n+\xi x_1(n-\bar n)$, where $\bar n=L/x_1$ and the factor $\xi$ is,
\begin{eqnarray}
	\xi=1-\frac{1}{\cosh(\sqrt{\frac{\Delta n^2\gamma}{\bar{n}\kappa^-}})+\sqrt{\frac{\gamma}{\bar{n}\kappa^-}}\sinh(\sqrt{\frac{\Delta n^2\gamma}{\bar{n}\kappa^-}})}\,.
	\label{eq:slope variation1}
\end{eqnarray}
Substituting $x_n$ into Eqs. (\ref{kappan}) and (\ref{pn}), the approximate expression of $n$-dependent dissipation rate becomes
\begin{equation}
	\kappa_{n}^+ \approx \kappa_{\rm v}+\frac{\kappa_0}{1+\exp(4\xi(L-nx_1)/d')}\, ,
\end{equation}
which means that the effective coupling width is further increased to $d'/\xi$, ultimately increasing the steady-state photon number fluctuations.

In the limit case of $\gamma\ll\kappa^\pm$, the factor $\xi\sim 0$, so $\kappa_n^+\sim \kappa_{\rm v}+\kappa_0/2$ and no longer depends on $n$. This leads to a thermal equilibrium steady state of the optical mode.

%\nocite{*}

\bibliography{references}

\begin{thebibliography}{90}%
\makeatletter
\providecommand \@ifxundefined [1]{%
 \@ifx{#1\undefined}
}%
\providecommand \@ifnum [1]{%
 \ifnum #1\expandafter \@firstoftwo
 \else \expandafter \@secondoftwo
 \fi
}%
\providecommand \@ifx [1]{%
 \ifx #1\expandafter \@firstoftwo
 \else \expandafter \@secondoftwo
 \fi
}%
\providecommand \natexlab [1]{#1}%
\providecommand \enquote  [1]{``#1''}%
\providecommand \bibnamefont  [1]{#1}%
\providecommand \bibfnamefont [1]{#1}%
\providecommand \citenamefont [1]{#1}%
\providecommand \href@noop [0]{\@secondoftwo}%
\providecommand \href [0]{\begingroup \@sanitize@url \@href}%
\providecommand \@href[1]{\@@startlink{#1}\@@href}%
\providecommand \@@href[1]{\endgroup#1\@@endlink}%
\providecommand \@sanitize@url [0]{\catcode `\\12\catcode `\$12\catcode
  `\&12\catcode `\#12\catcode `\^12\catcode `\_12\catcode `\%12\relax}%
\providecommand \@@startlink[1]{}%
\providecommand \@@endlink[0]{}%
\providecommand \url  [0]{\begingroup\@sanitize@url \@url }%
\providecommand \@url [1]{\endgroup\@href {#1}{\urlprefix }}%
\providecommand \urlprefix  [0]{URL }%
\providecommand \Eprint [0]{\href }%
\providecommand \doibase [0]{https://doi.org/}%
\providecommand \selectlanguage [0]{\@gobble}%
\providecommand \bibinfo  [0]{\@secondoftwo}%
\providecommand \bibfield  [0]{\@secondoftwo}%
\providecommand \translation [1]{[#1]}%
\providecommand \BibitemOpen [0]{}%
\providecommand \bibitemStop [0]{}%
\providecommand \bibitemNoStop [0]{.\EOS\space}%
\providecommand \EOS [0]{\spacefactor3000\relax}%
\providecommand \BibitemShut  [1]{\csname bibitem#1\endcsname}%
\let\auto@bib@innerbib\@empty
%</preamble>
\bibitem [{\citenamefont {You}\ \emph {et~al.}(2021)\citenamefont {You},
  \citenamefont {Hong}, \citenamefont {Bierhorst}, \citenamefont {Lita},
  \citenamefont {Glancy}, \citenamefont {Kolthammer}, \citenamefont {Knill},
  \citenamefont {Nam}, \citenamefont {Mirin}, \citenamefont
  {Maga{\~n}a-Loaiza},\ and\ \citenamefont {Gerrits}}]{you2021scalable}%
  \BibitemOpen
  \bibfield  {author} {\bibinfo {author} {\bibfnamefont {C.}~\bibnamefont
  {You}}, \bibinfo {author} {\bibfnamefont {M.}~\bibnamefont {Hong}}, \bibinfo
  {author} {\bibfnamefont {P.}~\bibnamefont {Bierhorst}}, \bibinfo {author}
  {\bibfnamefont {A.~E.}\ \bibnamefont {Lita}}, \bibinfo {author}
  {\bibfnamefont {S.}~\bibnamefont {Glancy}}, \bibinfo {author} {\bibfnamefont
  {S.}~\bibnamefont {Kolthammer}}, \bibinfo {author} {\bibfnamefont
  {E.}~\bibnamefont {Knill}}, \bibinfo {author} {\bibfnamefont {S.~W.}\
  \bibnamefont {Nam}}, \bibinfo {author} {\bibfnamefont {R.~P.}\ \bibnamefont
  {Mirin}}, \bibinfo {author} {\bibfnamefont {O.~S.}\ \bibnamefont
  {Maga{\~n}a-Loaiza}},\ and\ \bibinfo {author} {\bibfnamefont
  {T.}~\bibnamefont {Gerrits}},\ }\bibfield  {title} {\bibinfo {title}
  {Scalable multiphoton quantum metrology with neither pre-nor post-selected
  measurements},\ }\href@noop {} {\bibfield  {journal} {\bibinfo  {journal}
  {Applied Physics Reviews}\ }\textbf {\bibinfo {volume} {8}},\ \bibinfo
  {pages} {041406} (\bibinfo {year} {2021})}\BibitemShut {NoStop}%
\bibitem [{\citenamefont {Thekkadath}\ \emph {et~al.}(2020)\citenamefont
  {Thekkadath}, \citenamefont {Mycroft}, \citenamefont {Bell}, \citenamefont
  {Wade}, \citenamefont {Eckstein}, \citenamefont {Phillips}, \citenamefont
  {Patel}, \citenamefont {Buraczewski}, \citenamefont {Lita}, \citenamefont
  {Gerrits}, \citenamefont {Nam}, \citenamefont {Stobi\'nska}, \citenamefont
  {Lvovsky},\ and\ \citenamefont {Walmsley}}]{thekkadath2020quantum}%
  \BibitemOpen
  \bibfield  {author} {\bibinfo {author} {\bibfnamefont {G.}~\bibnamefont
  {Thekkadath}}, \bibinfo {author} {\bibfnamefont {M.}~\bibnamefont {Mycroft}},
  \bibinfo {author} {\bibfnamefont {B.}~\bibnamefont {Bell}}, \bibinfo {author}
  {\bibfnamefont {C.}~\bibnamefont {Wade}}, \bibinfo {author} {\bibfnamefont
  {A.}~\bibnamefont {Eckstein}}, \bibinfo {author} {\bibfnamefont
  {D.}~\bibnamefont {Phillips}}, \bibinfo {author} {\bibfnamefont
  {R.}~\bibnamefont {Patel}}, \bibinfo {author} {\bibfnamefont
  {A.}~\bibnamefont {Buraczewski}}, \bibinfo {author} {\bibfnamefont
  {A.}~\bibnamefont {Lita}}, \bibinfo {author} {\bibfnamefont {T.}~\bibnamefont
  {Gerrits}}, \bibinfo {author} {\bibfnamefont {S.}~\bibnamefont {Nam}},
  \bibinfo {author} {\bibfnamefont {M.}~\bibnamefont {Stobi\'nska}}, \bibinfo
  {author} {\bibfnamefont {A.}~\bibnamefont {Lvovsky}},\ and\ \bibinfo {author}
  {\bibfnamefont {I.}~\bibnamefont {Walmsley}},\ }\bibfield  {title} {\bibinfo
  {title} {Quantum-enhanced interferometry with large heralded photon-number
  states},\ }\href@noop {} {\bibfield  {journal} {\bibinfo  {journal} {NPJ
  quantum information}\ }\textbf {\bibinfo {volume} {6}},\ \bibinfo {pages}
  {89} (\bibinfo {year} {2020})}\BibitemShut {NoStop}%
\bibitem [{\citenamefont {Eaton}\ \emph {et~al.}(2023)\citenamefont {Eaton},
  \citenamefont {Hossameldin}, \citenamefont {Birrittella}, \citenamefont
  {Alsing}, \citenamefont {Gerry}, \citenamefont {Dong}, \citenamefont
  {Cuevas},\ and\ \citenamefont {Pfister}}]{eaton2023resolution}%
  \BibitemOpen
  \bibfield  {author} {\bibinfo {author} {\bibfnamefont {M.}~\bibnamefont
  {Eaton}}, \bibinfo {author} {\bibfnamefont {A.}~\bibnamefont {Hossameldin}},
  \bibinfo {author} {\bibfnamefont {R.~J.}\ \bibnamefont {Birrittella}},
  \bibinfo {author} {\bibfnamefont {P.~M.}\ \bibnamefont {Alsing}}, \bibinfo
  {author} {\bibfnamefont {C.~C.}\ \bibnamefont {Gerry}}, \bibinfo {author}
  {\bibfnamefont {H.}~\bibnamefont {Dong}}, \bibinfo {author} {\bibfnamefont
  {C.}~\bibnamefont {Cuevas}},\ and\ \bibinfo {author} {\bibfnamefont
  {O.}~\bibnamefont {Pfister}},\ }\bibfield  {title} {\bibinfo {title}
  {Resolution of 100 photons and quantum generation of unbiased random
  numbers},\ }\href@noop {} {\bibfield  {journal} {\bibinfo  {journal} {Nature
  Photonics}\ }\textbf {\bibinfo {volume} {17}},\ \bibinfo {pages} {106}
  (\bibinfo {year} {2023})}\BibitemShut {NoStop}%
\bibitem [{\citenamefont {Ansari}\ \emph {et~al.}(2018)\citenamefont {Ansari},
  \citenamefont {Donohue}, \citenamefont {Brecht},\ and\ \citenamefont
  {Silberhorn}}]{ansari2018tailoring}%
  \BibitemOpen
  \bibfield  {author} {\bibinfo {author} {\bibfnamefont {V.}~\bibnamefont
  {Ansari}}, \bibinfo {author} {\bibfnamefont {J.~M.}\ \bibnamefont {Donohue}},
  \bibinfo {author} {\bibfnamefont {B.}~\bibnamefont {Brecht}},\ and\ \bibinfo
  {author} {\bibfnamefont {C.}~\bibnamefont {Silberhorn}},\ }\bibfield  {title}
  {\bibinfo {title} {Tailoring nonlinear processes for quantum optics with
  pulsed temporal-mode encodings},\ }\href@noop {} {\bibfield  {journal}
  {\bibinfo  {journal} {Optica}\ }\textbf {\bibinfo {volume} {5}},\ \bibinfo
  {pages} {534} (\bibinfo {year} {2018})}\BibitemShut {NoStop}%
\bibitem [{\citenamefont {Madsen}\ \emph {et~al.}(2022)\citenamefont {Madsen},
  \citenamefont {Laudenbach}, \citenamefont {Askarani}, \citenamefont
  {Rortais}, \citenamefont {Vincent}, \citenamefont {Bulmer}, \citenamefont
  {Miatto}, \citenamefont {Neuhaus}, \citenamefont {Helt}, \citenamefont
  {Collins}, \citenamefont {Lita}, \citenamefont {Gerrits}, \citenamefont
  {Nam}, \citenamefont {Vaidya}, \citenamefont {Menotti}, \citenamefont
  {Dhand}, \citenamefont {Vernon}, \citenamefont {Quesada},\ and\ \citenamefont
  {Lavoie}}]{madsen2022quantum}%
  \BibitemOpen
  \bibfield  {author} {\bibinfo {author} {\bibfnamefont {L.~S.}\ \bibnamefont
  {Madsen}}, \bibinfo {author} {\bibfnamefont {F.}~\bibnamefont {Laudenbach}},
  \bibinfo {author} {\bibfnamefont {M.~F.}\ \bibnamefont {Askarani}}, \bibinfo
  {author} {\bibfnamefont {F.}~\bibnamefont {Rortais}}, \bibinfo {author}
  {\bibfnamefont {T.}~\bibnamefont {Vincent}}, \bibinfo {author} {\bibfnamefont
  {J.~F.}\ \bibnamefont {Bulmer}}, \bibinfo {author} {\bibfnamefont {F.~M.}\
  \bibnamefont {Miatto}}, \bibinfo {author} {\bibfnamefont {L.}~\bibnamefont
  {Neuhaus}}, \bibinfo {author} {\bibfnamefont {L.~G.}\ \bibnamefont {Helt}},
  \bibinfo {author} {\bibfnamefont {M.~J.}\ \bibnamefont {Collins}}, \bibinfo
  {author} {\bibfnamefont {A.~E.}\ \bibnamefont {Lita}}, \bibinfo {author}
  {\bibfnamefont {T.}~\bibnamefont {Gerrits}}, \bibinfo {author} {\bibfnamefont
  {S.~W.}\ \bibnamefont {Nam}}, \bibinfo {author} {\bibfnamefont {V.~D.}\
  \bibnamefont {Vaidya}}, \bibinfo {author} {\bibfnamefont {M.}~\bibnamefont
  {Menotti}}, \bibinfo {author} {\bibfnamefont {I.}~\bibnamefont {Dhand}},
  \bibinfo {author} {\bibfnamefont {Z.}~\bibnamefont {Vernon}}, \bibinfo
  {author} {\bibfnamefont {N.}~\bibnamefont {Quesada}},\ and\ \bibinfo {author}
  {\bibfnamefont {J.}~\bibnamefont {Lavoie}},\ }\bibfield  {title} {\bibinfo
  {title} {Quantum computational advantage with a programmable photonic
  processor},\ }\href@noop {} {\bibfield  {journal} {\bibinfo  {journal}
  {Nature}\ }\textbf {\bibinfo {volume} {606}},\ \bibinfo {pages} {75}
  (\bibinfo {year} {2022})}\BibitemShut {NoStop}%
\bibitem [{\citenamefont {Goldberg}\ and\ \citenamefont
  {Steinberg}(2020)}]{goldberg2020transcoherent}%
  \BibitemOpen
  \bibfield  {author} {\bibinfo {author} {\bibfnamefont {A.~Z.}\ \bibnamefont
  {Goldberg}}\ and\ \bibinfo {author} {\bibfnamefont {A.~M.}\ \bibnamefont
  {Steinberg}},\ }\bibfield  {title} {\bibinfo {title} {Transcoherent states:
  optical states for maximal generation of atomic coherence},\ }\href@noop {}
  {\bibfield  {journal} {\bibinfo  {journal} {PRX Quantum}\ }\textbf {\bibinfo
  {volume} {1}},\ \bibinfo {pages} {020306} (\bibinfo {year}
  {2020})}\BibitemShut {NoStop}%
\bibitem [{\citenamefont {Goldberg}\ \emph {et~al.}(2023)\citenamefont
  {Goldberg}, \citenamefont {Steinberg},\ and\ \citenamefont
  {Heshami}}]{goldberg2023beyond}%
  \BibitemOpen
  \bibfield  {author} {\bibinfo {author} {\bibfnamefont {A.~Z.}\ \bibnamefont
  {Goldberg}}, \bibinfo {author} {\bibfnamefont {A.~M.}\ \bibnamefont
  {Steinberg}},\ and\ \bibinfo {author} {\bibfnamefont {K.}~\bibnamefont
  {Heshami}},\ }\bibfield  {title} {\bibinfo {title} {Beyond transcoherent
  states: Field states for effecting optimal coherent rotations on single or
  multiple qubits},\ }\href@noop {} {\bibfield  {journal} {\bibinfo  {journal}
  {Quantum}\ }\textbf {\bibinfo {volume} {7}},\ \bibinfo {pages} {963}
  (\bibinfo {year} {2023})}\BibitemShut {NoStop}%
\bibitem [{\citenamefont {del Campo}\ and\ \citenamefont
  {Muga}(2008)}]{PhysRevA.78.023412}%
  \BibitemOpen
  \bibfield  {author} {\bibinfo {author} {\bibfnamefont {A.}~\bibnamefont {del
  Campo}}\ and\ \bibinfo {author} {\bibfnamefont {J.~G.}\ \bibnamefont
  {Muga}},\ }\bibfield  {title} {\bibinfo {title} {Atom fock-state preparation
  by trap reduction},\ }\href {https://doi.org/10.1103/PhysRevA.78.023412}
  {\bibfield  {journal} {\bibinfo  {journal} {Phys. Rev. A}\ }\textbf {\bibinfo
  {volume} {78}},\ \bibinfo {pages} {023412} (\bibinfo {year}
  {2008})}\BibitemShut {NoStop}%
\bibitem [{\citenamefont {Ebert}\ \emph {et~al.}(2014)\citenamefont {Ebert},
  \citenamefont {Gill}, \citenamefont {Gibbons}, \citenamefont {Zhang},
  \citenamefont {Saffman},\ and\ \citenamefont
  {Walker}}]{PhysRevLett.112.043602}%
  \BibitemOpen
  \bibfield  {author} {\bibinfo {author} {\bibfnamefont {M.}~\bibnamefont
  {Ebert}}, \bibinfo {author} {\bibfnamefont {A.}~\bibnamefont {Gill}},
  \bibinfo {author} {\bibfnamefont {M.}~\bibnamefont {Gibbons}}, \bibinfo
  {author} {\bibfnamefont {X.}~\bibnamefont {Zhang}}, \bibinfo {author}
  {\bibfnamefont {M.}~\bibnamefont {Saffman}},\ and\ \bibinfo {author}
  {\bibfnamefont {T.~G.}\ \bibnamefont {Walker}},\ }\bibfield  {title}
  {\bibinfo {title} {Atomic fock state preparation using rydberg blockade},\
  }\href {https://doi.org/10.1103/PhysRevLett.112.043602} {\bibfield  {journal}
  {\bibinfo  {journal} {Phys. Rev. Lett.}\ }\textbf {\bibinfo {volume} {112}},\
  \bibinfo {pages} {043602} (\bibinfo {year} {2014})}\BibitemShut {NoStop}%
\bibitem [{\citenamefont {D'Ariano}\ \emph {et~al.}(2000)\citenamefont
  {D'Ariano}, \citenamefont {Maccone}, \citenamefont {Paris},\ and\
  \citenamefont {Sacchi}}]{PhysRevA.61.053817}%
  \BibitemOpen
  \bibfield  {author} {\bibinfo {author} {\bibfnamefont {G.~M.}\ \bibnamefont
  {D'Ariano}}, \bibinfo {author} {\bibfnamefont {L.}~\bibnamefont {Maccone}},
  \bibinfo {author} {\bibfnamefont {M.~G.~A.}\ \bibnamefont {Paris}},\ and\
  \bibinfo {author} {\bibfnamefont {M.~F.}\ \bibnamefont {Sacchi}},\ }\bibfield
   {title} {\bibinfo {title} {Optical fock-state synthesizer},\ }\href
  {https://doi.org/10.1103/PhysRevA.61.053817} {\bibfield  {journal} {\bibinfo
  {journal} {Phys. Rev. A}\ }\textbf {\bibinfo {volume} {61}},\ \bibinfo
  {pages} {053817} (\bibinfo {year} {2000})}\BibitemShut {NoStop}%
\bibitem [{\citenamefont {Greif}\ \emph {et~al.}(2016)\citenamefont {Greif},
  \citenamefont {Parsons}, \citenamefont {Mazurenko}, \citenamefont {Chiu},
  \citenamefont {Blatt}, \citenamefont {Huber}, \citenamefont {Ji},\ and\
  \citenamefont {Greiner}}]{greif2016site}%
  \BibitemOpen
  \bibfield  {author} {\bibinfo {author} {\bibfnamefont {D.}~\bibnamefont
  {Greif}}, \bibinfo {author} {\bibfnamefont {M.~F.}\ \bibnamefont {Parsons}},
  \bibinfo {author} {\bibfnamefont {A.}~\bibnamefont {Mazurenko}}, \bibinfo
  {author} {\bibfnamefont {C.~S.}\ \bibnamefont {Chiu}}, \bibinfo {author}
  {\bibfnamefont {S.}~\bibnamefont {Blatt}}, \bibinfo {author} {\bibfnamefont
  {F.}~\bibnamefont {Huber}}, \bibinfo {author} {\bibfnamefont
  {G.}~\bibnamefont {Ji}},\ and\ \bibinfo {author} {\bibfnamefont
  {M.}~\bibnamefont {Greiner}},\ }\bibfield  {title} {\bibinfo {title}
  {Site-resolved imaging of a fermionic mott insulator},\ }\href@noop {}
  {\bibfield  {journal} {\bibinfo  {journal} {Science}\ }\textbf {\bibinfo
  {volume} {351}},\ \bibinfo {pages} {953} (\bibinfo {year}
  {2016})}\BibitemShut {NoStop}%
\bibitem [{\citenamefont {Serwane}\ \emph {et~al.}(2011)\citenamefont
  {Serwane}, \citenamefont {Z{\"u}rn}, \citenamefont {Lompe}, \citenamefont
  {Ottenstein}, \citenamefont {Wenz},\ and\ \citenamefont
  {Jochim}}]{serwane2011deterministic}%
  \BibitemOpen
  \bibfield  {author} {\bibinfo {author} {\bibfnamefont {F.}~\bibnamefont
  {Serwane}}, \bibinfo {author} {\bibfnamefont {G.}~\bibnamefont {Z{\"u}rn}},
  \bibinfo {author} {\bibfnamefont {T.}~\bibnamefont {Lompe}}, \bibinfo
  {author} {\bibfnamefont {T.}~\bibnamefont {Ottenstein}}, \bibinfo {author}
  {\bibfnamefont {A.}~\bibnamefont {Wenz}},\ and\ \bibinfo {author}
  {\bibfnamefont {S.}~\bibnamefont {Jochim}},\ }\bibfield  {title} {\bibinfo
  {title} {Deterministic preparation of a tunable few-fermion system},\
  }\href@noop {} {\bibfield  {journal} {\bibinfo  {journal} {Science}\ }\textbf
  {\bibinfo {volume} {332}},\ \bibinfo {pages} {336} (\bibinfo {year}
  {2011})}\BibitemShut {NoStop}%
\bibitem [{\citenamefont {Majumdar}\ \emph {et~al.}(2012)\citenamefont
  {Majumdar}, \citenamefont {Bajcsy}, \citenamefont {Rundquist},\ and\
  \citenamefont {Vu\ifmmode \check{c}\else
  \v{c}\fi{}kovi\ifmmode~\acute{c}\else \'{c}\fi{}}}]{PhysRevLett.108.183601}%
  \BibitemOpen
  \bibfield  {author} {\bibinfo {author} {\bibfnamefont {A.}~\bibnamefont
  {Majumdar}}, \bibinfo {author} {\bibfnamefont {M.}~\bibnamefont {Bajcsy}},
  \bibinfo {author} {\bibfnamefont {A.}~\bibnamefont {Rundquist}},\ and\
  \bibinfo {author} {\bibfnamefont {J.}~\bibnamefont {Vu\ifmmode \check{c}\else
  \v{c}\fi{}kovi\ifmmode~\acute{c}\else \'{c}\fi{}}},\ }\bibfield  {title}
  {\bibinfo {title} {Loss-enabled sub-poissonian light generation in a bimodal
  nanocavity},\ }\href {https://doi.org/10.1103/PhysRevLett.108.183601}
  {\bibfield  {journal} {\bibinfo  {journal} {Phys. Rev. Lett.}\ }\textbf
  {\bibinfo {volume} {108}},\ \bibinfo {pages} {183601} (\bibinfo {year}
  {2012})}\BibitemShut {NoStop}%
\bibitem [{\citenamefont {Flayac}\ and\ \citenamefont
  {Savona}(2017)}]{PhysRevA.96.053810}%
  \BibitemOpen
  \bibfield  {author} {\bibinfo {author} {\bibfnamefont {H.}~\bibnamefont
  {Flayac}}\ and\ \bibinfo {author} {\bibfnamefont {V.}~\bibnamefont
  {Savona}},\ }\bibfield  {title} {\bibinfo {title} {Unconventional photon
  blockade},\ }\href {https://doi.org/10.1103/PhysRevA.96.053810} {\bibfield
  {journal} {\bibinfo  {journal} {Phys. Rev. A}\ }\textbf {\bibinfo {volume}
  {96}},\ \bibinfo {pages} {053810} (\bibinfo {year} {2017})}\BibitemShut
  {NoStop}%
\bibitem [{\citenamefont {Lang}\ \emph {et~al.}(2011)\citenamefont {Lang},
  \citenamefont {Bozyigit}, \citenamefont {Eichler}, \citenamefont {Steffen},
  \citenamefont {Fink}, \citenamefont {Abdumalikov}, \citenamefont {Baur},
  \citenamefont {Filipp}, \citenamefont {da~Silva}, \citenamefont {Blais},\
  and\ \citenamefont {Wallraff}}]{PhysRevLett.106.243601}%
  \BibitemOpen
  \bibfield  {author} {\bibinfo {author} {\bibfnamefont {C.}~\bibnamefont
  {Lang}}, \bibinfo {author} {\bibfnamefont {D.}~\bibnamefont {Bozyigit}},
  \bibinfo {author} {\bibfnamefont {C.}~\bibnamefont {Eichler}}, \bibinfo
  {author} {\bibfnamefont {L.}~\bibnamefont {Steffen}}, \bibinfo {author}
  {\bibfnamefont {J.~M.}\ \bibnamefont {Fink}}, \bibinfo {author}
  {\bibfnamefont {A.~A.}\ \bibnamefont {Abdumalikov}}, \bibinfo {author}
  {\bibfnamefont {M.}~\bibnamefont {Baur}}, \bibinfo {author} {\bibfnamefont
  {S.}~\bibnamefont {Filipp}}, \bibinfo {author} {\bibfnamefont {M.~P.}\
  \bibnamefont {da~Silva}}, \bibinfo {author} {\bibfnamefont {A.}~\bibnamefont
  {Blais}},\ and\ \bibinfo {author} {\bibfnamefont {A.}~\bibnamefont
  {Wallraff}},\ }\bibfield  {title} {\bibinfo {title} {Observation of resonant
  photon blockade at microwave frequencies using correlation function
  measurements},\ }\href {https://doi.org/10.1103/PhysRevLett.106.243601}
  {\bibfield  {journal} {\bibinfo  {journal} {Phys. Rev. Lett.}\ }\textbf
  {\bibinfo {volume} {106}},\ \bibinfo {pages} {243601} (\bibinfo {year}
  {2011})}\BibitemShut {NoStop}%
\bibitem [{\citenamefont {Rabl}(2011)}]{PhysRevLett.107.063601}%
  \BibitemOpen
  \bibfield  {author} {\bibinfo {author} {\bibfnamefont {P.}~\bibnamefont
  {Rabl}},\ }\bibfield  {title} {\bibinfo {title} {Photon blockade effect in
  optomechanical systems},\ }\href
  {https://doi.org/10.1103/PhysRevLett.107.063601} {\bibfield  {journal}
  {\bibinfo  {journal} {Phys. Rev. Lett.}\ }\textbf {\bibinfo {volume} {107}},\
  \bibinfo {pages} {063601} (\bibinfo {year} {2011})}\BibitemShut {NoStop}%
\bibitem [{\citenamefont {Gevorgyan}\ \emph {et~al.}(2012)\citenamefont
  {Gevorgyan}, \citenamefont {Shahinyan},\ and\ \citenamefont
  {Kryuchkyan}}]{PhysRevA.85.053802}%
  \BibitemOpen
  \bibfield  {author} {\bibinfo {author} {\bibfnamefont {T.~V.}\ \bibnamefont
  {Gevorgyan}}, \bibinfo {author} {\bibfnamefont {A.~R.}\ \bibnamefont
  {Shahinyan}},\ and\ \bibinfo {author} {\bibfnamefont {G.~Y.}\ \bibnamefont
  {Kryuchkyan}},\ }\bibfield  {title} {\bibinfo {title} {Generation of fock
  states and qubits in periodically pulsed nonlinear oscillators},\ }\href
  {https://doi.org/10.1103/PhysRevA.85.053802} {\bibfield  {journal} {\bibinfo
  {journal} {Phys. Rev. A}\ }\textbf {\bibinfo {volume} {85}},\ \bibinfo
  {pages} {053802} (\bibinfo {year} {2012})}\BibitemShut {NoStop}%
\bibitem [{\citenamefont {Yanagimoto}\ \emph {et~al.}(2019)\citenamefont
  {Yanagimoto}, \citenamefont {Ng}, \citenamefont {Onodera},\ and\
  \citenamefont {Mabuchi}}]{PhysRevA.100.033822}%
  \BibitemOpen
  \bibfield  {author} {\bibinfo {author} {\bibfnamefont {R.}~\bibnamefont
  {Yanagimoto}}, \bibinfo {author} {\bibfnamefont {E.}~\bibnamefont {Ng}},
  \bibinfo {author} {\bibfnamefont {T.}~\bibnamefont {Onodera}},\ and\ \bibinfo
  {author} {\bibfnamefont {H.}~\bibnamefont {Mabuchi}},\ }\bibfield  {title}
  {\bibinfo {title} {Adiabatic fock-state-generation scheme using kerr
  nonlinearity},\ }\href {https://doi.org/10.1103/PhysRevA.100.033822}
  {\bibfield  {journal} {\bibinfo  {journal} {Phys. Rev. A}\ }\textbf {\bibinfo
  {volume} {100}},\ \bibinfo {pages} {033822} (\bibinfo {year}
  {2019})}\BibitemShut {NoStop}%
\bibitem [{\citenamefont {Wang}\ \emph {et~al.}(2017)\citenamefont {Wang},
  \citenamefont {Hu}, \citenamefont {Xu}, \citenamefont {Liu}, \citenamefont
  {Ma}, \citenamefont {Zheng}, \citenamefont {Vijay}, \citenamefont {Song},
  \citenamefont {Duan},\ and\ \citenamefont {Sun}}]{PhysRevLett.118.223604}%
  \BibitemOpen
  \bibfield  {author} {\bibinfo {author} {\bibfnamefont {W.}~\bibnamefont
  {Wang}}, \bibinfo {author} {\bibfnamefont {L.}~\bibnamefont {Hu}}, \bibinfo
  {author} {\bibfnamefont {Y.}~\bibnamefont {Xu}}, \bibinfo {author}
  {\bibfnamefont {K.}~\bibnamefont {Liu}}, \bibinfo {author} {\bibfnamefont
  {Y.}~\bibnamefont {Ma}}, \bibinfo {author} {\bibfnamefont {S.-B.}\
  \bibnamefont {Zheng}}, \bibinfo {author} {\bibfnamefont {R.}~\bibnamefont
  {Vijay}}, \bibinfo {author} {\bibfnamefont {Y.~P.}\ \bibnamefont {Song}},
  \bibinfo {author} {\bibfnamefont {L.-M.}\ \bibnamefont {Duan}},\ and\
  \bibinfo {author} {\bibfnamefont {L.}~\bibnamefont {Sun}},\ }\bibfield
  {title} {\bibinfo {title} {Converting quasiclassical states into arbitrary
  fock state superpositions in a superconducting circuit},\ }\href
  {https://doi.org/10.1103/PhysRevLett.118.223604} {\bibfield  {journal}
  {\bibinfo  {journal} {Phys. Rev. Lett.}\ }\textbf {\bibinfo {volume} {118}},\
  \bibinfo {pages} {223604} (\bibinfo {year} {2017})}\BibitemShut {NoStop}%
\bibitem [{\citenamefont {Lingenfelter}\ \emph {et~al.}(2021)\citenamefont
  {Lingenfelter}, \citenamefont {Roberts},\ and\ \citenamefont
  {Clerk}}]{doi:10.1126/sciadv.abj1916}%
  \BibitemOpen
  \bibfield  {author} {\bibinfo {author} {\bibfnamefont {A.}~\bibnamefont
  {Lingenfelter}}, \bibinfo {author} {\bibfnamefont {D.}~\bibnamefont
  {Roberts}},\ and\ \bibinfo {author} {\bibfnamefont {A.~A.}\ \bibnamefont
  {Clerk}},\ }\bibfield  {title} {\bibinfo {title} {Unconditional fock state
  generation using arbitrarily weak photonic nonlinearities},\ }\href
  {https://doi.org/10.1126/sciadv.abj1916} {\bibfield  {journal} {\bibinfo
  {journal} {Science Advances}\ }\textbf {\bibinfo {volume} {7}},\ \bibinfo
  {pages} {eabj1916} (\bibinfo {year} {2021})},\ \Eprint
  {https://arxiv.org/abs/https://www.science.org/doi/pdf/10.1126/sciadv.abj1916}
  {https://www.science.org/doi/pdf/10.1126/sciadv.abj1916} \BibitemShut
  {NoStop}%
\bibitem [{\citenamefont {Yamamoto}\ \emph {et~al.}(1992)\citenamefont
  {Yamamoto}, \citenamefont {Machida},\ and\ \citenamefont
  {Richardson}}]{yamamoto1992photon}%
  \BibitemOpen
  \bibfield  {author} {\bibinfo {author} {\bibfnamefont {Y.}~\bibnamefont
  {Yamamoto}}, \bibinfo {author} {\bibfnamefont {S.}~\bibnamefont {Machida}},\
  and\ \bibinfo {author} {\bibfnamefont {W.~H.}\ \bibnamefont {Richardson}},\
  }\bibfield  {title} {\bibinfo {title} {Photon number squeezed states in
  semiconductor lasers},\ }\href@noop {} {\bibfield  {journal} {\bibinfo
  {journal} {Science}\ }\textbf {\bibinfo {volume} {255}},\ \bibinfo {pages}
  {1219} (\bibinfo {year} {1992})}\BibitemShut {NoStop}%
\bibitem [{\citenamefont {Uria}\ \emph {et~al.}(2020)\citenamefont {Uria},
  \citenamefont {Solano},\ and\ \citenamefont
  {Hermann-Avigliano}}]{PhysRevLett.125.093603}%
  \BibitemOpen
  \bibfield  {author} {\bibinfo {author} {\bibfnamefont {M.}~\bibnamefont
  {Uria}}, \bibinfo {author} {\bibfnamefont {P.}~\bibnamefont {Solano}},\ and\
  \bibinfo {author} {\bibfnamefont {C.}~\bibnamefont {Hermann-Avigliano}},\
  }\bibfield  {title} {\bibinfo {title} {Deterministic generation of large fock
  states},\ }\href {https://doi.org/10.1103/PhysRevLett.125.093603} {\bibfield
  {journal} {\bibinfo  {journal} {Phys. Rev. Lett.}\ }\textbf {\bibinfo
  {volume} {125}},\ \bibinfo {pages} {093603} (\bibinfo {year}
  {2020})}\BibitemShut {NoStop}%
\bibitem [{\citenamefont {Meekhof}\ \emph {et~al.}(1996)\citenamefont
  {Meekhof}, \citenamefont {Monroe}, \citenamefont {King}, \citenamefont
  {Itano},\ and\ \citenamefont {Wineland}}]{PhysRevLett.76.1796}%
  \BibitemOpen
  \bibfield  {author} {\bibinfo {author} {\bibfnamefont {D.~M.}\ \bibnamefont
  {Meekhof}}, \bibinfo {author} {\bibfnamefont {C.}~\bibnamefont {Monroe}},
  \bibinfo {author} {\bibfnamefont {B.~E.}\ \bibnamefont {King}}, \bibinfo
  {author} {\bibfnamefont {W.~M.}\ \bibnamefont {Itano}},\ and\ \bibinfo
  {author} {\bibfnamefont {D.~J.}\ \bibnamefont {Wineland}},\ }\bibfield
  {title} {\bibinfo {title} {Generation of nonclassical motional states of a
  trapped atom},\ }\href {https://doi.org/10.1103/PhysRevLett.76.1796}
  {\bibfield  {journal} {\bibinfo  {journal} {Phys. Rev. Lett.}\ }\textbf
  {\bibinfo {volume} {76}},\ \bibinfo {pages} {1796} (\bibinfo {year}
  {1996})}\BibitemShut {NoStop}%
\bibitem [{\citenamefont {Krastanov}\ \emph {et~al.}(2015)\citenamefont
  {Krastanov}, \citenamefont {Albert}, \citenamefont {Shen}, \citenamefont
  {Zou}, \citenamefont {Heeres}, \citenamefont {Vlastakis}, \citenamefont
  {Schoelkopf},\ and\ \citenamefont {Jiang}}]{PhysRevA.92.040303}%
  \BibitemOpen
  \bibfield  {author} {\bibinfo {author} {\bibfnamefont {S.}~\bibnamefont
  {Krastanov}}, \bibinfo {author} {\bibfnamefont {V.~V.}\ \bibnamefont
  {Albert}}, \bibinfo {author} {\bibfnamefont {C.}~\bibnamefont {Shen}},
  \bibinfo {author} {\bibfnamefont {C.-L.}\ \bibnamefont {Zou}}, \bibinfo
  {author} {\bibfnamefont {R.~W.}\ \bibnamefont {Heeres}}, \bibinfo {author}
  {\bibfnamefont {B.}~\bibnamefont {Vlastakis}}, \bibinfo {author}
  {\bibfnamefont {R.~J.}\ \bibnamefont {Schoelkopf}},\ and\ \bibinfo {author}
  {\bibfnamefont {L.}~\bibnamefont {Jiang}},\ }\bibfield  {title} {\bibinfo
  {title} {Universal control of an oscillator with dispersive coupling to a
  qubit},\ }\href {https://doi.org/10.1103/PhysRevA.92.040303} {\bibfield
  {journal} {\bibinfo  {journal} {Phys. Rev. A}\ }\textbf {\bibinfo {volume}
  {92}},\ \bibinfo {pages} {040303} (\bibinfo {year} {2015})}\BibitemShut
  {NoStop}%
\bibitem [{\citenamefont {Parkins}\ \emph {et~al.}(1995)\citenamefont
  {Parkins}, \citenamefont {Marte}, \citenamefont {Zoller}, \citenamefont
  {Carnal},\ and\ \citenamefont {Kimble}}]{PhysRevA.51.1578}%
  \BibitemOpen
  \bibfield  {author} {\bibinfo {author} {\bibfnamefont {A.~S.}\ \bibnamefont
  {Parkins}}, \bibinfo {author} {\bibfnamefont {P.}~\bibnamefont {Marte}},
  \bibinfo {author} {\bibfnamefont {P.}~\bibnamefont {Zoller}}, \bibinfo
  {author} {\bibfnamefont {O.}~\bibnamefont {Carnal}},\ and\ \bibinfo {author}
  {\bibfnamefont {H.~J.}\ \bibnamefont {Kimble}},\ }\bibfield  {title}
  {\bibinfo {title} {Quantum-state mapping between multilevel atoms and cavity
  light fields},\ }\href {https://doi.org/10.1103/PhysRevA.51.1578} {\bibfield
  {journal} {\bibinfo  {journal} {Phys. Rev. A}\ }\textbf {\bibinfo {volume}
  {51}},\ \bibinfo {pages} {1578} (\bibinfo {year} {1995})}\BibitemShut
  {NoStop}%
\bibitem [{\citenamefont {xi~Liu}\ \emph {et~al.}(2004)\citenamefont {xi~Liu},
  \citenamefont {Wei},\ and\ \citenamefont {Nori}}]{Liu_2004}%
  \BibitemOpen
  \bibfield  {author} {\bibinfo {author} {\bibfnamefont {Y.}~\bibnamefont
  {xi~Liu}}, \bibinfo {author} {\bibfnamefont {L.~F.}\ \bibnamefont {Wei}},\
  and\ \bibinfo {author} {\bibfnamefont {F.}~\bibnamefont {Nori}},\ }\bibfield
  {title} {\bibinfo {title} {Generation of nonclassical photon states using a
  superconducting qubit in a microcavity},\ }\href
  {https://doi.org/10.1209/epl/i2004-10144-3} {\bibfield  {journal} {\bibinfo
  {journal} {Europhysics Letters ({EPL})}\ }\textbf {\bibinfo {volume} {67}},\
  \bibinfo {pages} {941} (\bibinfo {year} {2004})}\BibitemShut {NoStop}%
\bibitem [{\citenamefont {Heeres}\ \emph {et~al.}(2015)\citenamefont {Heeres},
  \citenamefont {Vlastakis}, \citenamefont {Holland}, \citenamefont
  {Krastanov}, \citenamefont {Albert}, \citenamefont {Frunzio}, \citenamefont
  {Jiang},\ and\ \citenamefont {Schoelkopf}}]{PhysRevLett.115.137002}%
  \BibitemOpen
  \bibfield  {author} {\bibinfo {author} {\bibfnamefont {R.~W.}\ \bibnamefont
  {Heeres}}, \bibinfo {author} {\bibfnamefont {B.}~\bibnamefont {Vlastakis}},
  \bibinfo {author} {\bibfnamefont {E.}~\bibnamefont {Holland}}, \bibinfo
  {author} {\bibfnamefont {S.}~\bibnamefont {Krastanov}}, \bibinfo {author}
  {\bibfnamefont {V.~V.}\ \bibnamefont {Albert}}, \bibinfo {author}
  {\bibfnamefont {L.}~\bibnamefont {Frunzio}}, \bibinfo {author} {\bibfnamefont
  {L.}~\bibnamefont {Jiang}},\ and\ \bibinfo {author} {\bibfnamefont {R.~J.}\
  \bibnamefont {Schoelkopf}},\ }\bibfield  {title} {\bibinfo {title} {Cavity
  state manipulation using photon-number selective phase gates},\ }\href
  {https://doi.org/10.1103/PhysRevLett.115.137002} {\bibfield  {journal}
  {\bibinfo  {journal} {Phys. Rev. Lett.}\ }\textbf {\bibinfo {volume} {115}},\
  \bibinfo {pages} {137002} (\bibinfo {year} {2015})}\BibitemShut {NoStop}%
\bibitem [{\citenamefont {Bertet}\ \emph {et~al.}(2002)\citenamefont {Bertet},
  \citenamefont {Osnaghi}, \citenamefont {Milman}, \citenamefont {Auffeves},
  \citenamefont {Maioli}, \citenamefont {Brune}, \citenamefont {Raimond},\ and\
  \citenamefont {Haroche}}]{PhysRevLett.88.143601}%
  \BibitemOpen
  \bibfield  {author} {\bibinfo {author} {\bibfnamefont {P.}~\bibnamefont
  {Bertet}}, \bibinfo {author} {\bibfnamefont {S.}~\bibnamefont {Osnaghi}},
  \bibinfo {author} {\bibfnamefont {P.}~\bibnamefont {Milman}}, \bibinfo
  {author} {\bibfnamefont {A.}~\bibnamefont {Auffeves}}, \bibinfo {author}
  {\bibfnamefont {P.}~\bibnamefont {Maioli}}, \bibinfo {author} {\bibfnamefont
  {M.}~\bibnamefont {Brune}}, \bibinfo {author} {\bibfnamefont {J.~M.}\
  \bibnamefont {Raimond}},\ and\ \bibinfo {author} {\bibfnamefont
  {S.}~\bibnamefont {Haroche}},\ }\bibfield  {title} {\bibinfo {title}
  {Generating and probing a two-photon fock state with a single atom in a
  cavity},\ }\href {https://doi.org/10.1103/PhysRevLett.88.143601} {\bibfield
  {journal} {\bibinfo  {journal} {Phys. Rev. Lett.}\ }\textbf {\bibinfo
  {volume} {88}},\ \bibinfo {pages} {143601} (\bibinfo {year}
  {2002})}\BibitemShut {NoStop}%
\bibitem [{\citenamefont {Weidinger}\ \emph {et~al.}(1999)\citenamefont
  {Weidinger}, \citenamefont {Varcoe}, \citenamefont {Heerlein},\ and\
  \citenamefont {Walther}}]{PhysRevLett.82.3795}%
  \BibitemOpen
  \bibfield  {author} {\bibinfo {author} {\bibfnamefont {M.}~\bibnamefont
  {Weidinger}}, \bibinfo {author} {\bibfnamefont {B.~T.~H.}\ \bibnamefont
  {Varcoe}}, \bibinfo {author} {\bibfnamefont {R.}~\bibnamefont {Heerlein}},\
  and\ \bibinfo {author} {\bibfnamefont {H.}~\bibnamefont {Walther}},\
  }\bibfield  {title} {\bibinfo {title} {Trapping states in the micromaser},\
  }\href {https://doi.org/10.1103/PhysRevLett.82.3795} {\bibfield  {journal}
  {\bibinfo  {journal} {Phys. Rev. Lett.}\ }\textbf {\bibinfo {volume} {82}},\
  \bibinfo {pages} {3795} (\bibinfo {year} {1999})}\BibitemShut {NoStop}%
\bibitem [{\citenamefont {Varcoe}\ \emph {et~al.}(2000)\citenamefont {Varcoe},
  \citenamefont {Brattke}, \citenamefont {Weidinger},\ and\ \citenamefont
  {Walther}}]{varcoe2000preparing}%
  \BibitemOpen
  \bibfield  {author} {\bibinfo {author} {\bibfnamefont {B.~T.}\ \bibnamefont
  {Varcoe}}, \bibinfo {author} {\bibfnamefont {S.}~\bibnamefont {Brattke}},
  \bibinfo {author} {\bibfnamefont {M.}~\bibnamefont {Weidinger}},\ and\
  \bibinfo {author} {\bibfnamefont {H.}~\bibnamefont {Walther}},\ }\bibfield
  {title} {\bibinfo {title} {Preparing pure photon number states of the
  radiation field},\ }\href@noop {} {\bibfield  {journal} {\bibinfo  {journal}
  {Nature}\ }\textbf {\bibinfo {volume} {403}},\ \bibinfo {pages} {743}
  (\bibinfo {year} {2000})}\BibitemShut {NoStop}%
\bibitem [{\citenamefont {Sayrin}\ \emph {et~al.}(2011)\citenamefont {Sayrin},
  \citenamefont {Dotsenko}, \citenamefont {Zhou}, \citenamefont {Peaudecerf},
  \citenamefont {Rybarczyk}, \citenamefont {Gleyzes}, \citenamefont {Rouchon},
  \citenamefont {Mirrahimi}, \citenamefont {Amini}, \citenamefont {Brune},
  \citenamefont {Raimond},\ and\ \citenamefont {Haroche}}]{sayrin2011real}%
  \BibitemOpen
  \bibfield  {author} {\bibinfo {author} {\bibfnamefont {C.}~\bibnamefont
  {Sayrin}}, \bibinfo {author} {\bibfnamefont {I.}~\bibnamefont {Dotsenko}},
  \bibinfo {author} {\bibfnamefont {X.}~\bibnamefont {Zhou}}, \bibinfo {author}
  {\bibfnamefont {B.}~\bibnamefont {Peaudecerf}}, \bibinfo {author}
  {\bibfnamefont {T.}~\bibnamefont {Rybarczyk}}, \bibinfo {author}
  {\bibfnamefont {S.}~\bibnamefont {Gleyzes}}, \bibinfo {author} {\bibfnamefont
  {P.}~\bibnamefont {Rouchon}}, \bibinfo {author} {\bibfnamefont
  {M.}~\bibnamefont {Mirrahimi}}, \bibinfo {author} {\bibfnamefont
  {H.}~\bibnamefont {Amini}}, \bibinfo {author} {\bibfnamefont
  {M.}~\bibnamefont {Brune}}, \bibinfo {author} {\bibfnamefont {J.-M.}\
  \bibnamefont {Raimond}},\ and\ \bibinfo {author} {\bibfnamefont
  {S.}~\bibnamefont {Haroche}},\ }\bibfield  {title} {\bibinfo {title}
  {Real-time quantum feedback prepares and stabilizes photon number states},\
  }\href@noop {} {\bibfield  {journal} {\bibinfo  {journal} {Nature}\ }\textbf
  {\bibinfo {volume} {477}},\ \bibinfo {pages} {73} (\bibinfo {year}
  {2011})}\BibitemShut {NoStop}%
\bibitem [{\citenamefont {Brown}\ \emph {et~al.}(2003)\citenamefont {Brown},
  \citenamefont {Dani}, \citenamefont {Stamper-Kurn},\ and\ \citenamefont
  {Whaley}}]{PhysRevA.67.043818}%
  \BibitemOpen
  \bibfield  {author} {\bibinfo {author} {\bibfnamefont {K.~R.}\ \bibnamefont
  {Brown}}, \bibinfo {author} {\bibfnamefont {K.~M.}\ \bibnamefont {Dani}},
  \bibinfo {author} {\bibfnamefont {D.~M.}\ \bibnamefont {Stamper-Kurn}},\ and\
  \bibinfo {author} {\bibfnamefont {K.~B.}\ \bibnamefont {Whaley}},\ }\bibfield
   {title} {\bibinfo {title} {Deterministic optical fock-state generation},\
  }\href {https://doi.org/10.1103/PhysRevA.67.043818} {\bibfield  {journal}
  {\bibinfo  {journal} {Phys. Rev. A}\ }\textbf {\bibinfo {volume} {67}},\
  \bibinfo {pages} {043818} (\bibinfo {year} {2003})}\BibitemShut {NoStop}%
\bibitem [{\citenamefont {Peaudecerf}\ \emph {et~al.}(2013)\citenamefont
  {Peaudecerf}, \citenamefont {Sayrin}, \citenamefont {Zhou}, \citenamefont
  {Rybarczyk}, \citenamefont {Gleyzes}, \citenamefont {Dotsenko}, \citenamefont
  {Raimond}, \citenamefont {Brune},\ and\ \citenamefont
  {Haroche}}]{PhysRevA.87.042320}%
  \BibitemOpen
  \bibfield  {author} {\bibinfo {author} {\bibfnamefont {B.}~\bibnamefont
  {Peaudecerf}}, \bibinfo {author} {\bibfnamefont {C.}~\bibnamefont {Sayrin}},
  \bibinfo {author} {\bibfnamefont {X.}~\bibnamefont {Zhou}}, \bibinfo {author}
  {\bibfnamefont {T.}~\bibnamefont {Rybarczyk}}, \bibinfo {author}
  {\bibfnamefont {S.}~\bibnamefont {Gleyzes}}, \bibinfo {author} {\bibfnamefont
  {I.}~\bibnamefont {Dotsenko}}, \bibinfo {author} {\bibfnamefont {J.~M.}\
  \bibnamefont {Raimond}}, \bibinfo {author} {\bibfnamefont {M.}~\bibnamefont
  {Brune}},\ and\ \bibinfo {author} {\bibfnamefont {S.}~\bibnamefont
  {Haroche}},\ }\bibfield  {title} {\bibinfo {title} {Quantum feedback
  experiments stabilizing fock states of light in a cavity},\ }\href
  {https://doi.org/10.1103/PhysRevA.87.042320} {\bibfield  {journal} {\bibinfo
  {journal} {Phys. Rev. A}\ }\textbf {\bibinfo {volume} {87}},\ \bibinfo
  {pages} {042320} (\bibinfo {year} {2013})}\BibitemShut {NoStop}%
\bibitem [{\citenamefont {Groiseau}\ \emph {et~al.}(2021)\citenamefont
  {Groiseau}, \citenamefont {Elliott}, \citenamefont {Masson},\ and\
  \citenamefont {Parkins}}]{PhysRevLett.127.033602}%
  \BibitemOpen
  \bibfield  {author} {\bibinfo {author} {\bibfnamefont {C.}~\bibnamefont
  {Groiseau}}, \bibinfo {author} {\bibfnamefont {A.~E.~J.}\ \bibnamefont
  {Elliott}}, \bibinfo {author} {\bibfnamefont {S.~J.}\ \bibnamefont
  {Masson}},\ and\ \bibinfo {author} {\bibfnamefont {S.}~\bibnamefont
  {Parkins}},\ }\bibfield  {title} {\bibinfo {title} {Proposal for a
  deterministic single-atom source of quasisuperradiant $n$-photon pulses},\
  }\href {https://doi.org/10.1103/PhysRevLett.127.033602} {\bibfield  {journal}
  {\bibinfo  {journal} {Phys. Rev. Lett.}\ }\textbf {\bibinfo {volume} {127}},\
  \bibinfo {pages} {033602} (\bibinfo {year} {2021})}\BibitemShut {NoStop}%
\bibitem [{\citenamefont {Fran\ifmmode \mbox{\c{c}}\else~\c{c}\fi{}a Santos}\
  \emph {et~al.}(2001)\citenamefont {Fran\ifmmode \mbox{\c{c}}\else~\c{c}\fi{}a
  Santos}, \citenamefont {Solano},\ and\ \citenamefont
  {de~Matos~Filho}}]{PhysRevLett.87.093601}%
  \BibitemOpen
  \bibfield  {author} {\bibinfo {author} {\bibfnamefont {M.}~\bibnamefont
  {Fran\ifmmode \mbox{\c{c}}\else~\c{c}\fi{}a Santos}}, \bibinfo {author}
  {\bibfnamefont {E.}~\bibnamefont {Solano}},\ and\ \bibinfo {author}
  {\bibfnamefont {R.~L.}\ \bibnamefont {de~Matos~Filho}},\ }\bibfield  {title}
  {\bibinfo {title} {Conditional large fock state preparation and field state
  reconstruction in cavity qed},\ }\href
  {https://doi.org/10.1103/PhysRevLett.87.093601} {\bibfield  {journal}
  {\bibinfo  {journal} {Phys. Rev. Lett.}\ }\textbf {\bibinfo {volume} {87}},\
  \bibinfo {pages} {093601} (\bibinfo {year} {2001})}\BibitemShut {NoStop}%
\bibitem [{\citenamefont {Hofheinz}\ \emph {et~al.}(2008)\citenamefont
  {Hofheinz}, \citenamefont {Weig}, \citenamefont {Ansmann}, \citenamefont
  {Bialczak}, \citenamefont {Lucero}, \citenamefont {Neeley}, \citenamefont
  {O’connell}, \citenamefont {Wang}, \citenamefont {Martinis},\ and\
  \citenamefont {Cleland}}]{hofheinz2008generation}%
  \BibitemOpen
  \bibfield  {author} {\bibinfo {author} {\bibfnamefont {M.}~\bibnamefont
  {Hofheinz}}, \bibinfo {author} {\bibfnamefont {E.}~\bibnamefont {Weig}},
  \bibinfo {author} {\bibfnamefont {M.}~\bibnamefont {Ansmann}}, \bibinfo
  {author} {\bibfnamefont {R.~C.}\ \bibnamefont {Bialczak}}, \bibinfo {author}
  {\bibfnamefont {E.}~\bibnamefont {Lucero}}, \bibinfo {author} {\bibfnamefont
  {M.}~\bibnamefont {Neeley}}, \bibinfo {author} {\bibfnamefont
  {A.}~\bibnamefont {O’connell}}, \bibinfo {author} {\bibfnamefont
  {H.}~\bibnamefont {Wang}}, \bibinfo {author} {\bibfnamefont {J.~M.}\
  \bibnamefont {Martinis}},\ and\ \bibinfo {author} {\bibfnamefont
  {A.}~\bibnamefont {Cleland}},\ }\bibfield  {title} {\bibinfo {title}
  {Generation of fock states in a superconducting quantum circuit},\
  }\href@noop {} {\bibfield  {journal} {\bibinfo  {journal} {Nature}\ }\textbf
  {\bibinfo {volume} {454}},\ \bibinfo {pages} {310} (\bibinfo {year}
  {2008})}\BibitemShut {NoStop}%
\bibitem [{\citenamefont {Premaratne}\ \emph {et~al.}(2017)\citenamefont
  {Premaratne}, \citenamefont {Wellstood},\ and\ \citenamefont
  {Palmer}}]{premaratne2017microwave}%
  \BibitemOpen
  \bibfield  {author} {\bibinfo {author} {\bibfnamefont {S.~P.}\ \bibnamefont
  {Premaratne}}, \bibinfo {author} {\bibfnamefont {F.}~\bibnamefont
  {Wellstood}},\ and\ \bibinfo {author} {\bibfnamefont {B.}~\bibnamefont
  {Palmer}},\ }\bibfield  {title} {\bibinfo {title} {Microwave photon fock
  state generation by stimulated raman adiabatic passage},\ }\href@noop {}
  {\bibfield  {journal} {\bibinfo  {journal} {Nature communications}\ }\textbf
  {\bibinfo {volume} {8}},\ \bibinfo {pages} {1} (\bibinfo {year}
  {2017})}\BibitemShut {NoStop}%
\bibitem [{\citenamefont {Dotsenko}\ \emph {et~al.}(2009)\citenamefont
  {Dotsenko}, \citenamefont {Mirrahimi}, \citenamefont {Brune}, \citenamefont
  {Haroche}, \citenamefont {Raimond},\ and\ \citenamefont
  {Rouchon}}]{PhysRevA.80.013805}%
  \BibitemOpen
  \bibfield  {author} {\bibinfo {author} {\bibfnamefont {I.}~\bibnamefont
  {Dotsenko}}, \bibinfo {author} {\bibfnamefont {M.}~\bibnamefont {Mirrahimi}},
  \bibinfo {author} {\bibfnamefont {M.}~\bibnamefont {Brune}}, \bibinfo
  {author} {\bibfnamefont {S.}~\bibnamefont {Haroche}}, \bibinfo {author}
  {\bibfnamefont {J.-M.}\ \bibnamefont {Raimond}},\ and\ \bibinfo {author}
  {\bibfnamefont {P.}~\bibnamefont {Rouchon}},\ }\bibfield  {title} {\bibinfo
  {title} {Quantum feedback by discrete quantum nondemolition measurements:
  Towards on-demand generation of photon-number states},\ }\href
  {https://doi.org/10.1103/PhysRevA.80.013805} {\bibfield  {journal} {\bibinfo
  {journal} {Phys. Rev. A}\ }\textbf {\bibinfo {volume} {80}},\ \bibinfo
  {pages} {013805} (\bibinfo {year} {2009})}\BibitemShut {NoStop}%
\bibitem [{\citenamefont {Parkins}\ \emph {et~al.}(1993)\citenamefont
  {Parkins}, \citenamefont {Marte}, \citenamefont {Zoller},\ and\ \citenamefont
  {Kimble}}]{PhysRevLett.71.3095}%
  \BibitemOpen
  \bibfield  {author} {\bibinfo {author} {\bibfnamefont {A.~S.}\ \bibnamefont
  {Parkins}}, \bibinfo {author} {\bibfnamefont {P.}~\bibnamefont {Marte}},
  \bibinfo {author} {\bibfnamefont {P.}~\bibnamefont {Zoller}},\ and\ \bibinfo
  {author} {\bibfnamefont {H.~J.}\ \bibnamefont {Kimble}},\ }\bibfield  {title}
  {\bibinfo {title} {Synthesis of arbitrary quantum states via adiabatic
  transfer of zeeman coherence},\ }\href
  {https://doi.org/10.1103/PhysRevLett.71.3095} {\bibfield  {journal} {\bibinfo
   {journal} {Phys. Rev. Lett.}\ }\textbf {\bibinfo {volume} {71}},\ \bibinfo
  {pages} {3095} (\bibinfo {year} {1993})}\BibitemShut {NoStop}%
\bibitem [{\citenamefont {Zhou}\ \emph {et~al.}(2012)\citenamefont {Zhou},
  \citenamefont {Dotsenko}, \citenamefont {Peaudecerf}, \citenamefont
  {Rybarczyk}, \citenamefont {Sayrin}, \citenamefont {Gleyzes}, \citenamefont
  {Raimond}, \citenamefont {Brune},\ and\ \citenamefont
  {Haroche}}]{PhysRevLett.108.243602}%
  \BibitemOpen
  \bibfield  {author} {\bibinfo {author} {\bibfnamefont {X.}~\bibnamefont
  {Zhou}}, \bibinfo {author} {\bibfnamefont {I.}~\bibnamefont {Dotsenko}},
  \bibinfo {author} {\bibfnamefont {B.}~\bibnamefont {Peaudecerf}}, \bibinfo
  {author} {\bibfnamefont {T.}~\bibnamefont {Rybarczyk}}, \bibinfo {author}
  {\bibfnamefont {C.}~\bibnamefont {Sayrin}}, \bibinfo {author} {\bibfnamefont
  {S.}~\bibnamefont {Gleyzes}}, \bibinfo {author} {\bibfnamefont {J.~M.}\
  \bibnamefont {Raimond}}, \bibinfo {author} {\bibfnamefont {M.}~\bibnamefont
  {Brune}},\ and\ \bibinfo {author} {\bibfnamefont {S.}~\bibnamefont
  {Haroche}},\ }\bibfield  {title} {\bibinfo {title} {Field locked to a fock
  state by quantum feedback with single photon corrections},\ }\href
  {https://doi.org/10.1103/PhysRevLett.108.243602} {\bibfield  {journal}
  {\bibinfo  {journal} {Phys. Rev. Lett.}\ }\textbf {\bibinfo {volume} {108}},\
  \bibinfo {pages} {243602} (\bibinfo {year} {2012})}\BibitemShut {NoStop}%
\bibitem [{\citenamefont {Hofheinz}\ \emph {et~al.}(2009)\citenamefont
  {Hofheinz}, \citenamefont {Wang}, \citenamefont {Ansmann}, \citenamefont
  {Bialczak}, \citenamefont {Lucero}, \citenamefont {Neeley}, \citenamefont
  {O'connell}, \citenamefont {Sank}, \citenamefont {Wenner}, \citenamefont
  {Martinis},\ and\ \citenamefont {Cleland}}]{hofheinz2009synthesizing}%
  \BibitemOpen
  \bibfield  {author} {\bibinfo {author} {\bibfnamefont {M.}~\bibnamefont
  {Hofheinz}}, \bibinfo {author} {\bibfnamefont {H.}~\bibnamefont {Wang}},
  \bibinfo {author} {\bibfnamefont {M.}~\bibnamefont {Ansmann}}, \bibinfo
  {author} {\bibfnamefont {R.~C.}\ \bibnamefont {Bialczak}}, \bibinfo {author}
  {\bibfnamefont {E.}~\bibnamefont {Lucero}}, \bibinfo {author} {\bibfnamefont
  {M.}~\bibnamefont {Neeley}}, \bibinfo {author} {\bibfnamefont
  {A.}~\bibnamefont {O'connell}}, \bibinfo {author} {\bibfnamefont
  {D.}~\bibnamefont {Sank}}, \bibinfo {author} {\bibfnamefont {J.}~\bibnamefont
  {Wenner}}, \bibinfo {author} {\bibfnamefont {J.~M.}\ \bibnamefont
  {Martinis}},\ and\ \bibinfo {author} {\bibfnamefont {A.}~\bibnamefont
  {Cleland}},\ }\bibfield  {title} {\bibinfo {title} {Synthesizing arbitrary
  quantum states in a superconducting resonator},\ }\href@noop {} {\bibfield
  {journal} {\bibinfo  {journal} {Nature}\ }\textbf {\bibinfo {volume} {459}},\
  \bibinfo {pages} {546} (\bibinfo {year} {2009})}\BibitemShut {NoStop}%
\bibitem [{\citenamefont {Canela}\ and\ \citenamefont
  {Carmichael}(2020)}]{PhysRevLett.124.063604}%
  \BibitemOpen
  \bibfield  {author} {\bibinfo {author} {\bibfnamefont {V.~S.~C.}\
  \bibnamefont {Canela}}\ and\ \bibinfo {author} {\bibfnamefont {H.~J.}\
  \bibnamefont {Carmichael}},\ }\bibfield  {title} {\bibinfo {title} {Bright
  sub-poissonian light through intrinsic feedback and external control},\
  }\href {https://doi.org/10.1103/PhysRevLett.124.063604} {\bibfield  {journal}
  {\bibinfo  {journal} {Phys. Rev. Lett.}\ }\textbf {\bibinfo {volume} {124}},\
  \bibinfo {pages} {063604} (\bibinfo {year} {2020})}\BibitemShut {NoStop}%
\bibitem [{\citenamefont {Vogel}\ \emph {et~al.}(1993)\citenamefont {Vogel},
  \citenamefont {Akulin},\ and\ \citenamefont
  {Schleich}}]{PhysRevLett.71.1816}%
  \BibitemOpen
  \bibfield  {author} {\bibinfo {author} {\bibfnamefont {K.}~\bibnamefont
  {Vogel}}, \bibinfo {author} {\bibfnamefont {V.~M.}\ \bibnamefont {Akulin}},\
  and\ \bibinfo {author} {\bibfnamefont {W.~P.}\ \bibnamefont {Schleich}},\
  }\bibfield  {title} {\bibinfo {title} {Quantum state engineering of the
  radiation field},\ }\href {https://doi.org/10.1103/PhysRevLett.71.1816}
  {\bibfield  {journal} {\bibinfo  {journal} {Phys. Rev. Lett.}\ }\textbf
  {\bibinfo {volume} {71}},\ \bibinfo {pages} {1816} (\bibinfo {year}
  {1993})}\BibitemShut {NoStop}%
\bibitem [{\citenamefont {Geremia}(2006)}]{PhysRevLett.97.073601}%
  \BibitemOpen
  \bibfield  {author} {\bibinfo {author} {\bibfnamefont {J.}~\bibnamefont
  {Geremia}},\ }\bibfield  {title} {\bibinfo {title} {Deterministic and
  nondestructively verifiable preparation of photon number states},\ }\href
  {https://doi.org/10.1103/PhysRevLett.97.073601} {\bibfield  {journal}
  {\bibinfo  {journal} {Phys. Rev. Lett.}\ }\textbf {\bibinfo {volume} {97}},\
  \bibinfo {pages} {073601} (\bibinfo {year} {2006})}\BibitemShut {NoStop}%
\bibitem [{\citenamefont {Hong}\ and\ \citenamefont
  {Mandel}(1986)}]{PhysRevLett.56.58}%
  \BibitemOpen
  \bibfield  {author} {\bibinfo {author} {\bibfnamefont {C.~K.}\ \bibnamefont
  {Hong}}\ and\ \bibinfo {author} {\bibfnamefont {L.}~\bibnamefont {Mandel}},\
  }\bibfield  {title} {\bibinfo {title} {Experimental realization of a
  localized one-photon state},\ }\href
  {https://doi.org/10.1103/PhysRevLett.56.58} {\bibfield  {journal} {\bibinfo
  {journal} {Phys. Rev. Lett.}\ }\textbf {\bibinfo {volume} {56}},\ \bibinfo
  {pages} {58} (\bibinfo {year} {1986})}\BibitemShut {NoStop}%
\bibitem [{\citenamefont {Guerlin}\ \emph {et~al.}(2007)\citenamefont
  {Guerlin}, \citenamefont {Bernu}, \citenamefont {Deleglise}, \citenamefont
  {Sayrin}, \citenamefont {Gleyzes}, \citenamefont {Kuhr}, \citenamefont
  {Brune}, \citenamefont {Raimond},\ and\ \citenamefont
  {Haroche}}]{guerlin2007progressive}%
  \BibitemOpen
  \bibfield  {author} {\bibinfo {author} {\bibfnamefont {C.}~\bibnamefont
  {Guerlin}}, \bibinfo {author} {\bibfnamefont {J.}~\bibnamefont {Bernu}},
  \bibinfo {author} {\bibfnamefont {S.}~\bibnamefont {Deleglise}}, \bibinfo
  {author} {\bibfnamefont {C.}~\bibnamefont {Sayrin}}, \bibinfo {author}
  {\bibfnamefont {S.}~\bibnamefont {Gleyzes}}, \bibinfo {author} {\bibfnamefont
  {S.}~\bibnamefont {Kuhr}}, \bibinfo {author} {\bibfnamefont {M.}~\bibnamefont
  {Brune}}, \bibinfo {author} {\bibfnamefont {J.-M.}\ \bibnamefont {Raimond}},\
  and\ \bibinfo {author} {\bibfnamefont {S.}~\bibnamefont {Haroche}},\
  }\bibfield  {title} {\bibinfo {title} {Progressive field-state collapse and
  quantum non-demolition photon counting},\ }\href@noop {} {\bibfield
  {journal} {\bibinfo  {journal} {Nature}\ }\textbf {\bibinfo {volume} {448}},\
  \bibinfo {pages} {889} (\bibinfo {year} {2007})}\BibitemShut {NoStop}%
\bibitem [{\citenamefont {Tiedau}\ \emph {et~al.}(2019)\citenamefont {Tiedau},
  \citenamefont {Bartley}, \citenamefont {Harder}, \citenamefont {Lita},
  \citenamefont {Nam}, \citenamefont {Gerrits},\ and\ \citenamefont
  {Silberhorn}}]{PhysRevA.100.041802}%
  \BibitemOpen
  \bibfield  {author} {\bibinfo {author} {\bibfnamefont {J.}~\bibnamefont
  {Tiedau}}, \bibinfo {author} {\bibfnamefont {T.~J.}\ \bibnamefont {Bartley}},
  \bibinfo {author} {\bibfnamefont {G.}~\bibnamefont {Harder}}, \bibinfo
  {author} {\bibfnamefont {A.~E.}\ \bibnamefont {Lita}}, \bibinfo {author}
  {\bibfnamefont {S.~W.}\ \bibnamefont {Nam}}, \bibinfo {author} {\bibfnamefont
  {T.}~\bibnamefont {Gerrits}},\ and\ \bibinfo {author} {\bibfnamefont
  {C.}~\bibnamefont {Silberhorn}},\ }\bibfield  {title} {\bibinfo {title}
  {Scalability of parametric down-conversion for generating higher-order fock
  states},\ }\href {https://doi.org/10.1103/PhysRevA.100.041802} {\bibfield
  {journal} {\bibinfo  {journal} {Phys. Rev. A}\ }\textbf {\bibinfo {volume}
  {100}},\ \bibinfo {pages} {041802} (\bibinfo {year} {2019})}\BibitemShut
  {NoStop}%
\bibitem [{\citenamefont {Harel}\ and\ \citenamefont
  {Kurizki}(1996)}]{PhysRevA.54.5410}%
  \BibitemOpen
  \bibfield  {author} {\bibinfo {author} {\bibfnamefont {G.}~\bibnamefont
  {Harel}}\ and\ \bibinfo {author} {\bibfnamefont {G.}~\bibnamefont
  {Kurizki}},\ }\bibfield  {title} {\bibinfo {title} {Fock-state preparation
  from thermal cavity fields by measurements on resonant atoms},\ }\href
  {https://doi.org/10.1103/PhysRevA.54.5410} {\bibfield  {journal} {\bibinfo
  {journal} {Phys. Rev. A}\ }\textbf {\bibinfo {volume} {54}},\ \bibinfo
  {pages} {5410} (\bibinfo {year} {1996})}\BibitemShut {NoStop}%
\bibitem [{\citenamefont {Cirac}\ \emph {et~al.}(1993)\citenamefont {Cirac},
  \citenamefont {Blatt}, \citenamefont {Parkins},\ and\ \citenamefont
  {Zoller}}]{PhysRevLett.70.762}%
  \BibitemOpen
  \bibfield  {author} {\bibinfo {author} {\bibfnamefont {J.~I.}\ \bibnamefont
  {Cirac}}, \bibinfo {author} {\bibfnamefont {R.}~\bibnamefont {Blatt}},
  \bibinfo {author} {\bibfnamefont {A.~S.}\ \bibnamefont {Parkins}},\ and\
  \bibinfo {author} {\bibfnamefont {P.}~\bibnamefont {Zoller}},\ }\bibfield
  {title} {\bibinfo {title} {Preparation of fock states by observation of
  quantum jumps in an ion trap},\ }\href
  {https://doi.org/10.1103/PhysRevLett.70.762} {\bibfield  {journal} {\bibinfo
  {journal} {Phys. Rev. Lett.}\ }\textbf {\bibinfo {volume} {70}},\ \bibinfo
  {pages} {762} (\bibinfo {year} {1993})}\BibitemShut {NoStop}%
\bibitem [{\citenamefont {Waks}\ \emph {et~al.}(2006)\citenamefont {Waks},
  \citenamefont {Diamanti},\ and\ \citenamefont {Yamamoto}}]{Waks_2006}%
  \BibitemOpen
  \bibfield  {author} {\bibinfo {author} {\bibfnamefont {E.}~\bibnamefont
  {Waks}}, \bibinfo {author} {\bibfnamefont {E.}~\bibnamefont {Diamanti}},\
  and\ \bibinfo {author} {\bibfnamefont {Y.}~\bibnamefont {Yamamoto}},\
  }\bibfield  {title} {\bibinfo {title} {Generation of photon number states},\
  }\href {https://doi.org/10.1088/1367-2630/8/1/004} {\bibfield  {journal}
  {\bibinfo  {journal} {New Journal of Physics}\ }\textbf {\bibinfo {volume}
  {8}},\ \bibinfo {pages} {4} (\bibinfo {year} {2006})}\BibitemShut {NoStop}%
\bibitem [{\citenamefont {Harder}\ \emph {et~al.}(2016)\citenamefont {Harder},
  \citenamefont {Bartley}, \citenamefont {Lita}, \citenamefont {Nam},
  \citenamefont {Gerrits},\ and\ \citenamefont
  {Silberhorn}}]{harder2016single}%
  \BibitemOpen
  \bibfield  {author} {\bibinfo {author} {\bibfnamefont {G.}~\bibnamefont
  {Harder}}, \bibinfo {author} {\bibfnamefont {T.~J.}\ \bibnamefont {Bartley}},
  \bibinfo {author} {\bibfnamefont {A.~E.}\ \bibnamefont {Lita}}, \bibinfo
  {author} {\bibfnamefont {S.~W.}\ \bibnamefont {Nam}}, \bibinfo {author}
  {\bibfnamefont {T.}~\bibnamefont {Gerrits}},\ and\ \bibinfo {author}
  {\bibfnamefont {C.}~\bibnamefont {Silberhorn}},\ }\bibfield  {title}
  {\bibinfo {title} {Single-mode parametric-down-conversion states with 50
  photons as a source for mesoscopic quantum optics},\ }\href@noop {}
  {\bibfield  {journal} {\bibinfo  {journal} {Physical review letters}\
  }\textbf {\bibinfo {volume} {116}},\ \bibinfo {pages} {143601} (\bibinfo
  {year} {2016})}\BibitemShut {NoStop}%
\bibitem [{\citenamefont {Poyatos}\ \emph {et~al.}(1996)\citenamefont
  {Poyatos}, \citenamefont {Cirac},\ and\ \citenamefont
  {Zoller}}]{PhysRevLett.77.4728}%
  \BibitemOpen
  \bibfield  {author} {\bibinfo {author} {\bibfnamefont {J.~F.}\ \bibnamefont
  {Poyatos}}, \bibinfo {author} {\bibfnamefont {J.~I.}\ \bibnamefont {Cirac}},\
  and\ \bibinfo {author} {\bibfnamefont {P.}~\bibnamefont {Zoller}},\
  }\bibfield  {title} {\bibinfo {title} {Quantum reservoir engineering with
  laser cooled trapped ions},\ }\href
  {https://doi.org/10.1103/PhysRevLett.77.4728} {\bibfield  {journal} {\bibinfo
   {journal} {Phys. Rev. Lett.}\ }\textbf {\bibinfo {volume} {77}},\ \bibinfo
  {pages} {4728} (\bibinfo {year} {1996})}\BibitemShut {NoStop}%
\bibitem [{\citenamefont {Diehl}\ \emph {et~al.}(2008)\citenamefont {Diehl},
  \citenamefont {Micheli}, \citenamefont {Kantian}, \citenamefont {Kraus},
  \citenamefont {B{\"u}chler},\ and\ \citenamefont
  {Zoller}}]{diehl2008quantum}%
  \BibitemOpen
  \bibfield  {author} {\bibinfo {author} {\bibfnamefont {S.}~\bibnamefont
  {Diehl}}, \bibinfo {author} {\bibfnamefont {A.}~\bibnamefont {Micheli}},
  \bibinfo {author} {\bibfnamefont {A.}~\bibnamefont {Kantian}}, \bibinfo
  {author} {\bibfnamefont {B.}~\bibnamefont {Kraus}}, \bibinfo {author}
  {\bibfnamefont {H.}~\bibnamefont {B{\"u}chler}},\ and\ \bibinfo {author}
  {\bibfnamefont {P.}~\bibnamefont {Zoller}},\ }\bibfield  {title} {\bibinfo
  {title} {Quantum states and phases in driven open quantum systems with cold
  atoms},\ }\href@noop {} {\bibfield  {journal} {\bibinfo  {journal} {Nature
  Physics}\ }\textbf {\bibinfo {volume} {4}},\ \bibinfo {pages} {878} (\bibinfo
  {year} {2008})}\BibitemShut {NoStop}%
\bibitem [{\citenamefont {Kraus}\ \emph {et~al.}(2008)\citenamefont {Kraus},
  \citenamefont {B{\"u}chler}, \citenamefont {Diehl}, \citenamefont {Kantian},
  \citenamefont {Micheli},\ and\ \citenamefont
  {Zoller}}]{kraus2008preparation}%
  \BibitemOpen
  \bibfield  {author} {\bibinfo {author} {\bibfnamefont {B.}~\bibnamefont
  {Kraus}}, \bibinfo {author} {\bibfnamefont {H.~P.}\ \bibnamefont
  {B{\"u}chler}}, \bibinfo {author} {\bibfnamefont {S.}~\bibnamefont {Diehl}},
  \bibinfo {author} {\bibfnamefont {A.}~\bibnamefont {Kantian}}, \bibinfo
  {author} {\bibfnamefont {A.}~\bibnamefont {Micheli}},\ and\ \bibinfo {author}
  {\bibfnamefont {P.}~\bibnamefont {Zoller}},\ }\bibfield  {title} {\bibinfo
  {title} {Preparation of entangled states by quantum markov processes},\
  }\href@noop {} {\bibfield  {journal} {\bibinfo  {journal} {Physical Review
  A}\ }\textbf {\bibinfo {volume} {78}},\ \bibinfo {pages} {042307} (\bibinfo
  {year} {2008})}\BibitemShut {NoStop}%
\bibitem [{\citenamefont {Muschik}\ \emph {et~al.}(2011)\citenamefont
  {Muschik}, \citenamefont {Polzik},\ and\ \citenamefont
  {Cirac}}]{muschik2011dissipatively}%
  \BibitemOpen
  \bibfield  {author} {\bibinfo {author} {\bibfnamefont {C.~A.}\ \bibnamefont
  {Muschik}}, \bibinfo {author} {\bibfnamefont {E.~S.}\ \bibnamefont
  {Polzik}},\ and\ \bibinfo {author} {\bibfnamefont {J.~I.}\ \bibnamefont
  {Cirac}},\ }\bibfield  {title} {\bibinfo {title} {Dissipatively driven
  entanglement of two macroscopic atomic ensembles},\ }\href@noop {} {\bibfield
   {journal} {\bibinfo  {journal} {Physical Review A}\ }\textbf {\bibinfo
  {volume} {83}},\ \bibinfo {pages} {052312} (\bibinfo {year}
  {2011})}\BibitemShut {NoStop}%
\bibitem [{\citenamefont {Lloyd}(2000)}]{PhysRevA.62.022108}%
  \BibitemOpen
  \bibfield  {author} {\bibinfo {author} {\bibfnamefont {S.}~\bibnamefont
  {Lloyd}},\ }\bibfield  {title} {\bibinfo {title} {Coherent quantum
  feedback},\ }\href {https://doi.org/10.1103/PhysRevA.62.022108} {\bibfield
  {journal} {\bibinfo  {journal} {Phys. Rev. A}\ }\textbf {\bibinfo {volume}
  {62}},\ \bibinfo {pages} {022108} (\bibinfo {year} {2000})}\BibitemShut
  {NoStop}%
\bibitem [{\citenamefont {de~Assis}\ \emph {et~al.}(2019)\citenamefont
  {de~Assis}, \citenamefont {de~Mendon\ifmmode~\mbox{\c{c}}\else \c{c}\fi{}a},
  \citenamefont {Villas-Boas}, \citenamefont {de~Souza}, \citenamefont
  {Sarthour}, \citenamefont {Oliveira},\ and\ \citenamefont
  {de~Almeida}}]{PhysRevLett.122.240602}%
  \BibitemOpen
  \bibfield  {author} {\bibinfo {author} {\bibfnamefont {R.~J.}\ \bibnamefont
  {de~Assis}}, \bibinfo {author} {\bibfnamefont {T.~M.}\ \bibnamefont
  {de~Mendon\ifmmode~\mbox{\c{c}}\else \c{c}\fi{}a}}, \bibinfo {author}
  {\bibfnamefont {C.~J.}\ \bibnamefont {Villas-Boas}}, \bibinfo {author}
  {\bibfnamefont {A.~M.}\ \bibnamefont {de~Souza}}, \bibinfo {author}
  {\bibfnamefont {R.~S.}\ \bibnamefont {Sarthour}}, \bibinfo {author}
  {\bibfnamefont {I.~S.}\ \bibnamefont {Oliveira}},\ and\ \bibinfo {author}
  {\bibfnamefont {N.~G.}\ \bibnamefont {de~Almeida}},\ }\bibfield  {title}
  {\bibinfo {title} {Efficiency of a quantum otto heat engine operating under a
  reservoir at effective negative temperatures},\ }\href
  {https://doi.org/10.1103/PhysRevLett.122.240602} {\bibfield  {journal}
  {\bibinfo  {journal} {Phys. Rev. Lett.}\ }\textbf {\bibinfo {volume} {122}},\
  \bibinfo {pages} {240602} (\bibinfo {year} {2019})}\BibitemShut {NoStop}%
\bibitem [{\citenamefont {Zhang}\ \emph {et~al.}(2013)\citenamefont {Zhang},
  \citenamefont {Meystre},\ and\ \citenamefont {Zhang}}]{PhysRevA.88.043632}%
  \BibitemOpen
  \bibfield  {author} {\bibinfo {author} {\bibfnamefont {K.}~\bibnamefont
  {Zhang}}, \bibinfo {author} {\bibfnamefont {P.}~\bibnamefont {Meystre}},\
  and\ \bibinfo {author} {\bibfnamefont {W.}~\bibnamefont {Zhang}},\ }\bibfield
   {title} {\bibinfo {title} {Back-action-free quantum optomechanics with
  negative-mass bose-einstein condensates},\ }\href
  {https://doi.org/10.1103/PhysRevA.88.043632} {\bibfield  {journal} {\bibinfo
  {journal} {Phys. Rev. A}\ }\textbf {\bibinfo {volume} {88}},\ \bibinfo
  {pages} {043632} (\bibinfo {year} {2013})}\BibitemShut {NoStop}%
\bibitem [{\citenamefont {Braun}\ \emph {et~al.}(2015)\citenamefont {Braun},
  \citenamefont {Friesdorf}, \citenamefont {Hodgman}, \citenamefont
  {Schreiber}, \citenamefont {Ronzheimer}, \citenamefont {Riera}, \citenamefont
  {del Rey}, \citenamefont {Bloch}, \citenamefont {Eisert},\ and\ \citenamefont
  {Schneider}}]{doi:10.1073/pnas.1408861112}%
  \BibitemOpen
  \bibfield  {author} {\bibinfo {author} {\bibfnamefont {S.}~\bibnamefont
  {Braun}}, \bibinfo {author} {\bibfnamefont {M.}~\bibnamefont {Friesdorf}},
  \bibinfo {author} {\bibfnamefont {S.~S.}\ \bibnamefont {Hodgman}}, \bibinfo
  {author} {\bibfnamefont {M.}~\bibnamefont {Schreiber}}, \bibinfo {author}
  {\bibfnamefont {J.~P.}\ \bibnamefont {Ronzheimer}}, \bibinfo {author}
  {\bibfnamefont {A.}~\bibnamefont {Riera}}, \bibinfo {author} {\bibfnamefont
  {M.}~\bibnamefont {del Rey}}, \bibinfo {author} {\bibfnamefont
  {I.}~\bibnamefont {Bloch}}, \bibinfo {author} {\bibfnamefont
  {J.}~\bibnamefont {Eisert}},\ and\ \bibinfo {author} {\bibfnamefont
  {U.}~\bibnamefont {Schneider}},\ }\bibfield  {title} {\bibinfo {title}
  {Emergence of coherence and the dynamics of quantum phase transitions},\
  }\href {https://doi.org/10.1073/pnas.1408861112} {\bibfield  {journal}
  {\bibinfo  {journal} {Proceedings of the National Academy of Sciences}\
  }\textbf {\bibinfo {volume} {112}},\ \bibinfo {pages} {3641} (\bibinfo {year}
  {2015})},\ \Eprint
  {https://arxiv.org/abs/https://www.pnas.org/doi/pdf/10.1073/pnas.1408861112}
  {https://www.pnas.org/doi/pdf/10.1073/pnas.1408861112} \BibitemShut {NoStop}%
\bibitem [{\citenamefont {Holland}\ \emph {et~al.}(2015)\citenamefont
  {Holland}, \citenamefont {Vlastakis}, \citenamefont {Heeres}, \citenamefont
  {Reagor}, \citenamefont {Vool}, \citenamefont {Leghtas}, \citenamefont
  {Frunzio}, \citenamefont {Kirchmair}, \citenamefont {Devoret}, \citenamefont
  {Mirrahimi},\ and\ \citenamefont {Schoelkopf}}]{PhysRevLett.115.180501}%
  \BibitemOpen
  \bibfield  {author} {\bibinfo {author} {\bibfnamefont {E.~T.}\ \bibnamefont
  {Holland}}, \bibinfo {author} {\bibfnamefont {B.}~\bibnamefont {Vlastakis}},
  \bibinfo {author} {\bibfnamefont {R.~W.}\ \bibnamefont {Heeres}}, \bibinfo
  {author} {\bibfnamefont {M.~J.}\ \bibnamefont {Reagor}}, \bibinfo {author}
  {\bibfnamefont {U.}~\bibnamefont {Vool}}, \bibinfo {author} {\bibfnamefont
  {Z.}~\bibnamefont {Leghtas}}, \bibinfo {author} {\bibfnamefont
  {L.}~\bibnamefont {Frunzio}}, \bibinfo {author} {\bibfnamefont
  {G.}~\bibnamefont {Kirchmair}}, \bibinfo {author} {\bibfnamefont {M.~H.}\
  \bibnamefont {Devoret}}, \bibinfo {author} {\bibfnamefont {M.}~\bibnamefont
  {Mirrahimi}},\ and\ \bibinfo {author} {\bibfnamefont {R.~J.}\ \bibnamefont
  {Schoelkopf}},\ }\bibfield  {title} {\bibinfo {title} {Single-photon-resolved
  cross-kerr interaction for autonomous stabilization of photon-number
  states},\ }\href {https://doi.org/10.1103/PhysRevLett.115.180501} {\bibfield
  {journal} {\bibinfo  {journal} {Phys. Rev. Lett.}\ }\textbf {\bibinfo
  {volume} {115}},\ \bibinfo {pages} {180501} (\bibinfo {year}
  {2015})}\BibitemShut {NoStop}%
\bibitem [{\citenamefont {Souquet}\ and\ \citenamefont
  {Clerk}(2016)}]{PhysRevA.93.060301}%
  \BibitemOpen
  \bibfield  {author} {\bibinfo {author} {\bibfnamefont {J.-R.}\ \bibnamefont
  {Souquet}}\ and\ \bibinfo {author} {\bibfnamefont {A.~A.}\ \bibnamefont
  {Clerk}},\ }\bibfield  {title} {\bibinfo {title} {Fock-state stabilization
  and emission in superconducting circuits using dc-biased josephson
  junctions},\ }\href {https://doi.org/10.1103/PhysRevA.93.060301} {\bibfield
  {journal} {\bibinfo  {journal} {Phys. Rev. A}\ }\textbf {\bibinfo {volume}
  {93}},\ \bibinfo {pages} {060301} (\bibinfo {year} {2016})}\BibitemShut
  {NoStop}%
\bibitem [{\citenamefont {Ben~Arosh}\ \emph {et~al.}(2021)\citenamefont
  {Ben~Arosh}, \citenamefont {Cross},\ and\ \citenamefont
  {Lifshitz}}]{PhysRevResearch.3.013130}%
  \BibitemOpen
  \bibfield  {author} {\bibinfo {author} {\bibfnamefont {L.}~\bibnamefont
  {Ben~Arosh}}, \bibinfo {author} {\bibfnamefont {M.~C.}\ \bibnamefont
  {Cross}},\ and\ \bibinfo {author} {\bibfnamefont {R.}~\bibnamefont
  {Lifshitz}},\ }\bibfield  {title} {\bibinfo {title} {Quantum limit cycles and
  the rayleigh and van der pol oscillators},\ }\href
  {https://doi.org/10.1103/PhysRevResearch.3.013130} {\bibfield  {journal}
  {\bibinfo  {journal} {Phys. Rev. Research}\ }\textbf {\bibinfo {volume}
  {3}},\ \bibinfo {pages} {013130} (\bibinfo {year} {2021})}\BibitemShut
  {NoStop}%
\bibitem [{\citenamefont {Kleckner}\ and\ \citenamefont
  {Bouwmeester}(2006)}]{kleckner2006sub}%
  \BibitemOpen
  \bibfield  {author} {\bibinfo {author} {\bibfnamefont {D.}~\bibnamefont
  {Kleckner}}\ and\ \bibinfo {author} {\bibfnamefont {D.}~\bibnamefont
  {Bouwmeester}},\ }\bibfield  {title} {\bibinfo {title} {Sub-kelvin optical
  cooling of a micromechanical resonator},\ }\href@noop {} {\bibfield
  {journal} {\bibinfo  {journal} {Nature}\ }\textbf {\bibinfo {volume} {444}},\
  \bibinfo {pages} {75} (\bibinfo {year} {2006})}\BibitemShut {NoStop}%
\bibitem [{\citenamefont {Li}\ and\ \citenamefont
  {Li}(2013)}]{li2013millikelvin}%
  \BibitemOpen
  \bibfield  {author} {\bibinfo {author} {\bibfnamefont {T.}~\bibnamefont
  {Li}}\ and\ \bibinfo {author} {\bibfnamefont {T.}~\bibnamefont {Li}},\
  }\bibfield  {title} {\bibinfo {title} {Millikelvin cooling of an optically
  trapped microsphere in vacuum},\ }\href@noop {} {\bibfield  {journal}
  {\bibinfo  {journal} {Fundamental Tests of Physics with Optically Trapped
  Microspheres}\ ,\ \bibinfo {pages} {81}} (\bibinfo {year}
  {2013})}\BibitemShut {NoStop}%
\bibitem [{\citenamefont {Magrini}\ \emph {et~al.}(2021)\citenamefont
  {Magrini}, \citenamefont {Rosenzweig}, \citenamefont {Bach}, \citenamefont
  {Deutschmann-Olek}, \citenamefont {Hofer}, \citenamefont {Hong},
  \citenamefont {Kiesel}, \citenamefont {Kugi},\ and\ \citenamefont
  {Aspelmeyer}}]{magrini2021real}%
  \BibitemOpen
  \bibfield  {author} {\bibinfo {author} {\bibfnamefont {L.}~\bibnamefont
  {Magrini}}, \bibinfo {author} {\bibfnamefont {P.}~\bibnamefont {Rosenzweig}},
  \bibinfo {author} {\bibfnamefont {C.}~\bibnamefont {Bach}}, \bibinfo {author}
  {\bibfnamefont {A.}~\bibnamefont {Deutschmann-Olek}}, \bibinfo {author}
  {\bibfnamefont {S.~G.}\ \bibnamefont {Hofer}}, \bibinfo {author}
  {\bibfnamefont {S.}~\bibnamefont {Hong}}, \bibinfo {author} {\bibfnamefont
  {N.}~\bibnamefont {Kiesel}}, \bibinfo {author} {\bibfnamefont
  {A.}~\bibnamefont {Kugi}},\ and\ \bibinfo {author} {\bibfnamefont
  {M.}~\bibnamefont {Aspelmeyer}},\ }\bibfield  {title} {\bibinfo {title}
  {Real-time optimal quantum control of mechanical motion at room
  temperature},\ }\href@noop {} {\bibfield  {journal} {\bibinfo  {journal}
  {Nature}\ }\textbf {\bibinfo {volume} {595}},\ \bibinfo {pages} {373}
  (\bibinfo {year} {2021})}\BibitemShut {NoStop}%
\bibitem [{\citenamefont {Tebbenjohanns}\ \emph {et~al.}(2021)\citenamefont
  {Tebbenjohanns}, \citenamefont {Mattana}, \citenamefont {Rossi},
  \citenamefont {Frimmer},\ and\ \citenamefont
  {Novotny}}]{tebbenjohanns2021quantum}%
  \BibitemOpen
  \bibfield  {author} {\bibinfo {author} {\bibfnamefont {F.}~\bibnamefont
  {Tebbenjohanns}}, \bibinfo {author} {\bibfnamefont {M.~L.}\ \bibnamefont
  {Mattana}}, \bibinfo {author} {\bibfnamefont {M.}~\bibnamefont {Rossi}},
  \bibinfo {author} {\bibfnamefont {M.}~\bibnamefont {Frimmer}},\ and\ \bibinfo
  {author} {\bibfnamefont {L.}~\bibnamefont {Novotny}},\ }\bibfield  {title}
  {\bibinfo {title} {Quantum control of a nanoparticle optically levitated in
  cryogenic free space},\ }\href@noop {} {\bibfield  {journal} {\bibinfo
  {journal} {Nature}\ }\textbf {\bibinfo {volume} {595}},\ \bibinfo {pages}
  {378} (\bibinfo {year} {2021})}\BibitemShut {NoStop}%
\bibitem [{\citenamefont {Elste}\ \emph {et~al.}(2009)\citenamefont {Elste},
  \citenamefont {Girvin},\ and\ \citenamefont
  {Clerk}}]{PhysRevLett.102.207209}%
  \BibitemOpen
  \bibfield  {author} {\bibinfo {author} {\bibfnamefont {F.}~\bibnamefont
  {Elste}}, \bibinfo {author} {\bibfnamefont {S.~M.}\ \bibnamefont {Girvin}},\
  and\ \bibinfo {author} {\bibfnamefont {A.~A.}\ \bibnamefont {Clerk}},\
  }\bibfield  {title} {\bibinfo {title} {Quantum noise interference and
  backaction cooling in cavity nanomechanics},\ }\href
  {https://doi.org/10.1103/PhysRevLett.102.207209} {\bibfield  {journal}
  {\bibinfo  {journal} {Phys. Rev. Lett.}\ }\textbf {\bibinfo {volume} {102}},\
  \bibinfo {pages} {207209} (\bibinfo {year} {2009})}\BibitemShut {NoStop}%
\bibitem [{\citenamefont {Chang}\ \emph {et~al.}(2014)\citenamefont {Chang},
  \citenamefont {Jiang}, \citenamefont {Hua}, \citenamefont {Yang},
  \citenamefont {Wen}, \citenamefont {Jiang}, \citenamefont {Li}, \citenamefont
  {Wang},\ and\ \citenamefont {Xiao}}]{chang2014parity}%
  \BibitemOpen
  \bibfield  {author} {\bibinfo {author} {\bibfnamefont {L.}~\bibnamefont
  {Chang}}, \bibinfo {author} {\bibfnamefont {X.}~\bibnamefont {Jiang}},
  \bibinfo {author} {\bibfnamefont {S.}~\bibnamefont {Hua}}, \bibinfo {author}
  {\bibfnamefont {C.}~\bibnamefont {Yang}}, \bibinfo {author} {\bibfnamefont
  {J.}~\bibnamefont {Wen}}, \bibinfo {author} {\bibfnamefont {L.}~\bibnamefont
  {Jiang}}, \bibinfo {author} {\bibfnamefont {G.}~\bibnamefont {Li}}, \bibinfo
  {author} {\bibfnamefont {G.}~\bibnamefont {Wang}},\ and\ \bibinfo {author}
  {\bibfnamefont {M.}~\bibnamefont {Xiao}},\ }\bibfield  {title} {\bibinfo
  {title} {Parity--time symmetry and variable optical isolation in
  active--passive-coupled microresonators},\ }\href@noop {} {\bibfield
  {journal} {\bibinfo  {journal} {Nature photonics}\ }\textbf {\bibinfo
  {volume} {8}},\ \bibinfo {pages} {524} (\bibinfo {year} {2014})}\BibitemShut
  {NoStop}%
\bibitem [{\citenamefont {Hodaei}\ \emph {et~al.}(2017)\citenamefont {Hodaei},
  \citenamefont {Hassan}, \citenamefont {Wittek}, \citenamefont
  {Garcia-Gracia}, \citenamefont {El-Ganainy}, \citenamefont
  {Christodoulides},\ and\ \citenamefont {Khajavikhan}}]{hodaei2017enhanced}%
  \BibitemOpen
  \bibfield  {author} {\bibinfo {author} {\bibfnamefont {H.}~\bibnamefont
  {Hodaei}}, \bibinfo {author} {\bibfnamefont {A.~U.}\ \bibnamefont {Hassan}},
  \bibinfo {author} {\bibfnamefont {S.}~\bibnamefont {Wittek}}, \bibinfo
  {author} {\bibfnamefont {H.}~\bibnamefont {Garcia-Gracia}}, \bibinfo {author}
  {\bibfnamefont {R.}~\bibnamefont {El-Ganainy}}, \bibinfo {author}
  {\bibfnamefont {D.~N.}\ \bibnamefont {Christodoulides}},\ and\ \bibinfo
  {author} {\bibfnamefont {M.}~\bibnamefont {Khajavikhan}},\ }\bibfield
  {title} {\bibinfo {title} {Enhanced sensitivity at higher-order exceptional
  points},\ }\href@noop {} {\bibfield  {journal} {\bibinfo  {journal} {Nature}\
  }\textbf {\bibinfo {volume} {548}},\ \bibinfo {pages} {187} (\bibinfo {year}
  {2017})}\BibitemShut {NoStop}%
\bibitem [{\citenamefont {El-Ganainy}\ \emph {et~al.}(2018)\citenamefont
  {El-Ganainy}, \citenamefont {Makris}, \citenamefont {Khajavikhan},
  \citenamefont {Musslimani}, \citenamefont {Rotter},\ and\ \citenamefont
  {Christodoulides}}]{el2018non}%
  \BibitemOpen
  \bibfield  {author} {\bibinfo {author} {\bibfnamefont {R.}~\bibnamefont
  {El-Ganainy}}, \bibinfo {author} {\bibfnamefont {K.~G.}\ \bibnamefont
  {Makris}}, \bibinfo {author} {\bibfnamefont {M.}~\bibnamefont {Khajavikhan}},
  \bibinfo {author} {\bibfnamefont {Z.~H.}\ \bibnamefont {Musslimani}},
  \bibinfo {author} {\bibfnamefont {S.}~\bibnamefont {Rotter}},\ and\ \bibinfo
  {author} {\bibfnamefont {D.~N.}\ \bibnamefont {Christodoulides}},\ }\bibfield
   {title} {\bibinfo {title} {Non-hermitian physics and pt symmetry},\
  }\href@noop {} {\bibfield  {journal} {\bibinfo  {journal} {Nature Physics}\
  }\textbf {\bibinfo {volume} {14}},\ \bibinfo {pages} {11} (\bibinfo {year}
  {2018})}\BibitemShut {NoStop}%
\bibitem [{\citenamefont {Peng}\ \emph {et~al.}(2014)\citenamefont {Peng},
  \citenamefont {{\"O}zdemir}, \citenamefont {Lei}, \citenamefont {Monifi},
  \citenamefont {Gianfreda}, \citenamefont {Long}, \citenamefont {Fan},
  \citenamefont {Nori}, \citenamefont {Bender},\ and\ \citenamefont
  {Yang}}]{peng2014parity}%
  \BibitemOpen
  \bibfield  {author} {\bibinfo {author} {\bibfnamefont {B.}~\bibnamefont
  {Peng}}, \bibinfo {author} {\bibfnamefont {{\c{S}}.~K.}\ \bibnamefont
  {{\"O}zdemir}}, \bibinfo {author} {\bibfnamefont {F.}~\bibnamefont {Lei}},
  \bibinfo {author} {\bibfnamefont {F.}~\bibnamefont {Monifi}}, \bibinfo
  {author} {\bibfnamefont {M.}~\bibnamefont {Gianfreda}}, \bibinfo {author}
  {\bibfnamefont {G.~L.}\ \bibnamefont {Long}}, \bibinfo {author}
  {\bibfnamefont {S.}~\bibnamefont {Fan}}, \bibinfo {author} {\bibfnamefont
  {F.}~\bibnamefont {Nori}}, \bibinfo {author} {\bibfnamefont {C.~M.}\
  \bibnamefont {Bender}},\ and\ \bibinfo {author} {\bibfnamefont
  {L.}~\bibnamefont {Yang}},\ }\bibfield  {title} {\bibinfo {title}
  {Parity--time-symmetric whispering-gallery microcavities},\ }\href@noop {}
  {\bibfield  {journal} {\bibinfo  {journal} {Nature Physics}\ }\textbf
  {\bibinfo {volume} {10}},\ \bibinfo {pages} {394} (\bibinfo {year}
  {2014})}\BibitemShut {NoStop}%
\bibitem [{\citenamefont {Marques~Muniz}\ \emph {et~al.}(2023)\citenamefont
  {Marques~Muniz}, \citenamefont {Wu}, \citenamefont {Jung}, \citenamefont
  {Khajavikhan}, \citenamefont {Christodoulides},\ and\ \citenamefont
  {Peschel}}]{marques2023observation}%
  \BibitemOpen
  \bibfield  {author} {\bibinfo {author} {\bibfnamefont {A.}~\bibnamefont
  {Marques~Muniz}}, \bibinfo {author} {\bibfnamefont {F.}~\bibnamefont {Wu}},
  \bibinfo {author} {\bibfnamefont {P.}~\bibnamefont {Jung}}, \bibinfo {author}
  {\bibfnamefont {M.}~\bibnamefont {Khajavikhan}}, \bibinfo {author}
  {\bibfnamefont {D.}~\bibnamefont {Christodoulides}},\ and\ \bibinfo {author}
  {\bibfnamefont {U.}~\bibnamefont {Peschel}},\ }\bibfield  {title} {\bibinfo
  {title} {Observation of photon-photon thermodynamic processes under negative
  optical temperature conditions},\ }\href@noop {} {\bibfield  {journal}
  {\bibinfo  {journal} {Science}\ }\textbf {\bibinfo {volume} {379}},\ \bibinfo
  {pages} {1019} (\bibinfo {year} {2023})}\BibitemShut {NoStop}%
\bibitem [{\citenamefont {Novotn\'y}\ \emph {et~al.}(2003)\citenamefont
  {Novotn\'y}, \citenamefont {Donarini},\ and\ \citenamefont
  {Jauho}}]{PhysRevLett.90.256801}%
  \BibitemOpen
  \bibfield  {author} {\bibinfo {author} {\bibfnamefont {T.~c.~v.}\
  \bibnamefont {Novotn\'y}}, \bibinfo {author} {\bibfnamefont {A.}~\bibnamefont
  {Donarini}},\ and\ \bibinfo {author} {\bibfnamefont {A.-P.}\ \bibnamefont
  {Jauho}},\ }\bibfield  {title} {\bibinfo {title} {Quantum shuttle in phase
  space},\ }\href {https://doi.org/10.1103/PhysRevLett.90.256801} {\bibfield
  {journal} {\bibinfo  {journal} {Phys. Rev. Lett.}\ }\textbf {\bibinfo
  {volume} {90}},\ \bibinfo {pages} {256801} (\bibinfo {year}
  {2003})}\BibitemShut {NoStop}%
\bibitem [{\citenamefont {Moskalenko}\ \emph {et~al.}(2009)\citenamefont
  {Moskalenko}, \citenamefont {Gordeev}, \citenamefont {Koentjoro},
  \citenamefont {Raithby}, \citenamefont {French}, \citenamefont {Marken},\
  and\ \citenamefont {Savel'ev}}]{PhysRevB.79.241403}%
  \BibitemOpen
  \bibfield  {author} {\bibinfo {author} {\bibfnamefont {A.~V.}\ \bibnamefont
  {Moskalenko}}, \bibinfo {author} {\bibfnamefont {S.~N.}\ \bibnamefont
  {Gordeev}}, \bibinfo {author} {\bibfnamefont {O.~F.}\ \bibnamefont
  {Koentjoro}}, \bibinfo {author} {\bibfnamefont {P.~R.}\ \bibnamefont
  {Raithby}}, \bibinfo {author} {\bibfnamefont {R.~W.}\ \bibnamefont {French}},
  \bibinfo {author} {\bibfnamefont {F.}~\bibnamefont {Marken}},\ and\ \bibinfo
  {author} {\bibfnamefont {S.~E.}\ \bibnamefont {Savel'ev}},\ }\bibfield
  {title} {\bibinfo {title} {Nanomechanical electron shuttle consisting of a
  gold nanoparticle embedded within the gap between two gold electrodes},\
  }\href {https://doi.org/10.1103/PhysRevB.79.241403} {\bibfield  {journal}
  {\bibinfo  {journal} {Phys. Rev. B}\ }\textbf {\bibinfo {volume} {79}},\
  \bibinfo {pages} {241403} (\bibinfo {year} {2009})}\BibitemShut {NoStop}%
\bibitem [{\citenamefont {Utami}\ \emph {et~al.}(2006)\citenamefont {Utami},
  \citenamefont {Goan}, \citenamefont {Holmes},\ and\ \citenamefont
  {Milburn}}]{PhysRevB.74.014303}%
  \BibitemOpen
  \bibfield  {author} {\bibinfo {author} {\bibfnamefont {D.~W.}\ \bibnamefont
  {Utami}}, \bibinfo {author} {\bibfnamefont {H.-S.}\ \bibnamefont {Goan}},
  \bibinfo {author} {\bibfnamefont {C.~A.}\ \bibnamefont {Holmes}},\ and\
  \bibinfo {author} {\bibfnamefont {G.~J.}\ \bibnamefont {Milburn}},\
  }\bibfield  {title} {\bibinfo {title} {Quantum noise in the electromechanical
  shuttle: Quantum master equation treatment},\ }\href
  {https://doi.org/10.1103/PhysRevB.74.014303} {\bibfield  {journal} {\bibinfo
  {journal} {Phys. Rev. B}\ }\textbf {\bibinfo {volume} {74}},\ \bibinfo
  {pages} {014303} (\bibinfo {year} {2006})}\BibitemShut {NoStop}%
\bibitem [{\citenamefont {Xuereb}\ \emph {et~al.}(2011)\citenamefont {Xuereb},
  \citenamefont {Schnabel},\ and\ \citenamefont
  {Hammerer}}]{PhysRevLett.107.213604}%
  \BibitemOpen
  \bibfield  {author} {\bibinfo {author} {\bibfnamefont {A.}~\bibnamefont
  {Xuereb}}, \bibinfo {author} {\bibfnamefont {R.}~\bibnamefont {Schnabel}},\
  and\ \bibinfo {author} {\bibfnamefont {K.}~\bibnamefont {Hammerer}},\
  }\bibfield  {title} {\bibinfo {title} {Dissipative optomechanics in a
  michelson-sagnac interferometer},\ }\href
  {https://doi.org/10.1103/PhysRevLett.107.213604} {\bibfield  {journal}
  {\bibinfo  {journal} {Phys. Rev. Lett.}\ }\textbf {\bibinfo {volume} {107}},\
  \bibinfo {pages} {213604} (\bibinfo {year} {2011})}\BibitemShut {NoStop}%
\bibitem [{\citenamefont {Tagantsev}\ and\ \citenamefont
  {Polzik}(2021)}]{tagantsev2021dissipative}%
  \BibitemOpen
  \bibfield  {author} {\bibinfo {author} {\bibfnamefont {A.~K.}\ \bibnamefont
  {Tagantsev}}\ and\ \bibinfo {author} {\bibfnamefont {E.~S.}\ \bibnamefont
  {Polzik}},\ }\bibfield  {title} {\bibinfo {title} {Dissipative optomechanical
  coupling with a membrane outside of an optical cavity},\ }\href@noop {}
  {\bibfield  {journal} {\bibinfo  {journal} {Physical Review A}\ }\textbf
  {\bibinfo {volume} {103}},\ \bibinfo {pages} {063503} (\bibinfo {year}
  {2021})}\BibitemShut {NoStop}%
\bibitem [{\citenamefont {Sankey}\ \emph {et~al.}(2010)\citenamefont {Sankey},
  \citenamefont {Yang}, \citenamefont {Zwickl}, \citenamefont {Jayich},\ and\
  \citenamefont {Harris}}]{sankey2010strong}%
  \BibitemOpen
  \bibfield  {author} {\bibinfo {author} {\bibfnamefont {J.~C.}\ \bibnamefont
  {Sankey}}, \bibinfo {author} {\bibfnamefont {C.}~\bibnamefont {Yang}},
  \bibinfo {author} {\bibfnamefont {B.~M.}\ \bibnamefont {Zwickl}}, \bibinfo
  {author} {\bibfnamefont {A.~M.}\ \bibnamefont {Jayich}},\ and\ \bibinfo
  {author} {\bibfnamefont {J.~G.}\ \bibnamefont {Harris}},\ }\bibfield  {title}
  {\bibinfo {title} {Strong and tunable nonlinear optomechanical coupling in a
  low-loss system},\ }\href@noop {} {\bibfield  {journal} {\bibinfo  {journal}
  {Nature Physics}\ }\textbf {\bibinfo {volume} {6}},\ \bibinfo {pages} {707}
  (\bibinfo {year} {2010})}\BibitemShut {NoStop}%
\bibitem [{\citenamefont {Kleckner}\ \emph {et~al.}(2011)\citenamefont
  {Kleckner}, \citenamefont {Pepper}, \citenamefont {Jeffrey}, \citenamefont
  {Sonin}, \citenamefont {Thon},\ and\ \citenamefont
  {Bouwmeester}}]{kleckner2011optomechanical}%
  \BibitemOpen
  \bibfield  {author} {\bibinfo {author} {\bibfnamefont {D.}~\bibnamefont
  {Kleckner}}, \bibinfo {author} {\bibfnamefont {B.}~\bibnamefont {Pepper}},
  \bibinfo {author} {\bibfnamefont {E.}~\bibnamefont {Jeffrey}}, \bibinfo
  {author} {\bibfnamefont {P.}~\bibnamefont {Sonin}}, \bibinfo {author}
  {\bibfnamefont {S.~M.}\ \bibnamefont {Thon}},\ and\ \bibinfo {author}
  {\bibfnamefont {D.}~\bibnamefont {Bouwmeester}},\ }\bibfield  {title}
  {\bibinfo {title} {Optomechanical trampoline resonators},\ }\href@noop {}
  {\bibfield  {journal} {\bibinfo  {journal} {Optics express}\ }\textbf
  {\bibinfo {volume} {19}},\ \bibinfo {pages} {19708} (\bibinfo {year}
  {2011})}\BibitemShut {NoStop}%
\bibitem [{\citenamefont {Favero}\ \emph {et~al.}(2007)\citenamefont {Favero},
  \citenamefont {Metzger}, \citenamefont {Camerer}, \citenamefont {K{\"o}nig},
  \citenamefont {Lorenz}, \citenamefont {Kotthaus},\ and\ \citenamefont
  {Karrai}}]{favero2007optical}%
  \BibitemOpen
  \bibfield  {author} {\bibinfo {author} {\bibfnamefont {I.}~\bibnamefont
  {Favero}}, \bibinfo {author} {\bibfnamefont {C.}~\bibnamefont {Metzger}},
  \bibinfo {author} {\bibfnamefont {S.}~\bibnamefont {Camerer}}, \bibinfo
  {author} {\bibfnamefont {D.}~\bibnamefont {K{\"o}nig}}, \bibinfo {author}
  {\bibfnamefont {H.}~\bibnamefont {Lorenz}}, \bibinfo {author} {\bibfnamefont
  {J.~P.}\ \bibnamefont {Kotthaus}},\ and\ \bibinfo {author} {\bibfnamefont
  {K.}~\bibnamefont {Karrai}},\ }\bibfield  {title} {\bibinfo {title} {Optical
  cooling of a micromirror of wavelength size},\ }\href@noop {} {\bibfield
  {journal} {\bibinfo  {journal} {Applied Physics Letters}\ }\textbf {\bibinfo
  {volume} {90}},\ \bibinfo {pages} {104101} (\bibinfo {year}
  {2007})}\BibitemShut {NoStop}%
\bibitem [{\citenamefont {Thompson}\ \emph {et~al.}(2008)\citenamefont
  {Thompson}, \citenamefont {Zwickl}, \citenamefont {Jayich}, \citenamefont
  {Marquardt}, \citenamefont {Girvin},\ and\ \citenamefont
  {Harris}}]{thompson2008strong}%
  \BibitemOpen
  \bibfield  {author} {\bibinfo {author} {\bibfnamefont {J.}~\bibnamefont
  {Thompson}}, \bibinfo {author} {\bibfnamefont {B.}~\bibnamefont {Zwickl}},
  \bibinfo {author} {\bibfnamefont {A.}~\bibnamefont {Jayich}}, \bibinfo
  {author} {\bibfnamefont {F.}~\bibnamefont {Marquardt}}, \bibinfo {author}
  {\bibfnamefont {S.}~\bibnamefont {Girvin}},\ and\ \bibinfo {author}
  {\bibfnamefont {J.}~\bibnamefont {Harris}},\ }\bibfield  {title} {\bibinfo
  {title} {Strong dispersive coupling of a high-finesse cavity to a
  micromechanical membrane},\ }\href@noop {} {\bibfield  {journal} {\bibinfo
  {journal} {Nature}\ }\textbf {\bibinfo {volume} {452}},\ \bibinfo {pages}
  {72} (\bibinfo {year} {2008})}\BibitemShut {NoStop}%
\bibitem [{\citenamefont {Li}\ \emph {et~al.}(2009)\citenamefont {Li},
  \citenamefont {Pernice},\ and\ \citenamefont
  {Tang}}]{PhysRevLett.103.223901}%
  \BibitemOpen
  \bibfield  {author} {\bibinfo {author} {\bibfnamefont {M.}~\bibnamefont
  {Li}}, \bibinfo {author} {\bibfnamefont {W.~H.~P.}\ \bibnamefont {Pernice}},\
  and\ \bibinfo {author} {\bibfnamefont {H.~X.}\ \bibnamefont {Tang}},\
  }\bibfield  {title} {\bibinfo {title} {Reactive cavity optical force on
  microdisk-coupled nanomechanical beam waveguides},\ }\href
  {https://doi.org/10.1103/PhysRevLett.103.223901} {\bibfield  {journal}
  {\bibinfo  {journal} {Phys. Rev. Lett.}\ }\textbf {\bibinfo {volume} {103}},\
  \bibinfo {pages} {223901} (\bibinfo {year} {2009})}\BibitemShut {NoStop}%
\bibitem [{\citenamefont {Carmon}\ \emph {et~al.}(2005)\citenamefont {Carmon},
  \citenamefont {Rokhsari}, \citenamefont {Yang}, \citenamefont {Kippenberg},\
  and\ \citenamefont {Vahala}}]{PhysRevLett.94.223902}%
  \BibitemOpen
  \bibfield  {author} {\bibinfo {author} {\bibfnamefont {T.}~\bibnamefont
  {Carmon}}, \bibinfo {author} {\bibfnamefont {H.}~\bibnamefont {Rokhsari}},
  \bibinfo {author} {\bibfnamefont {L.}~\bibnamefont {Yang}}, \bibinfo {author}
  {\bibfnamefont {T.~J.}\ \bibnamefont {Kippenberg}},\ and\ \bibinfo {author}
  {\bibfnamefont {K.~J.}\ \bibnamefont {Vahala}},\ }\bibfield  {title}
  {\bibinfo {title} {Temporal behavior of radiation-pressure-induced vibrations
  of an optical microcavity phonon mode},\ }\href
  {https://doi.org/10.1103/PhysRevLett.94.223902} {\bibfield  {journal}
  {\bibinfo  {journal} {Phys. Rev. Lett.}\ }\textbf {\bibinfo {volume} {94}},\
  \bibinfo {pages} {223902} (\bibinfo {year} {2005})}\BibitemShut {NoStop}%
\bibitem [{\citenamefont {Deli{\'c}}\ \emph {et~al.}(2020)\citenamefont
  {Deli{\'c}}, \citenamefont {Reisenbauer}, \citenamefont {Dare}, \citenamefont
  {Grass}, \citenamefont {Vuleti{\'c}}, \citenamefont {Kiesel},\ and\
  \citenamefont {Aspelmeyer}}]{delic2020cooling}%
  \BibitemOpen
  \bibfield  {author} {\bibinfo {author} {\bibfnamefont {U.}~\bibnamefont
  {Deli{\'c}}}, \bibinfo {author} {\bibfnamefont {M.}~\bibnamefont
  {Reisenbauer}}, \bibinfo {author} {\bibfnamefont {K.}~\bibnamefont {Dare}},
  \bibinfo {author} {\bibfnamefont {D.}~\bibnamefont {Grass}}, \bibinfo
  {author} {\bibfnamefont {V.}~\bibnamefont {Vuleti{\'c}}}, \bibinfo {author}
  {\bibfnamefont {N.}~\bibnamefont {Kiesel}},\ and\ \bibinfo {author}
  {\bibfnamefont {M.}~\bibnamefont {Aspelmeyer}},\ }\bibfield  {title}
  {\bibinfo {title} {Cooling of a levitated nanoparticle to the motional
  quantum ground state},\ }\href@noop {} {\bibfield  {journal} {\bibinfo
  {journal} {Science}\ }\textbf {\bibinfo {volume} {367}},\ \bibinfo {pages}
  {892} (\bibinfo {year} {2020})}\BibitemShut {NoStop}%
\bibitem [{\citenamefont {Ranfagni}\ \emph {et~al.}(2021)\citenamefont
  {Ranfagni}, \citenamefont {Vezio}, \citenamefont {Calamai}, \citenamefont
  {Chowdhury}, \citenamefont {Marino},\ and\ \citenamefont
  {Marin}}]{ranfagni2021vectorial}%
  \BibitemOpen
  \bibfield  {author} {\bibinfo {author} {\bibfnamefont {A.}~\bibnamefont
  {Ranfagni}}, \bibinfo {author} {\bibfnamefont {P.}~\bibnamefont {Vezio}},
  \bibinfo {author} {\bibfnamefont {M.}~\bibnamefont {Calamai}}, \bibinfo
  {author} {\bibfnamefont {A.}~\bibnamefont {Chowdhury}}, \bibinfo {author}
  {\bibfnamefont {F.}~\bibnamefont {Marino}},\ and\ \bibinfo {author}
  {\bibfnamefont {F.}~\bibnamefont {Marin}},\ }\bibfield  {title} {\bibinfo
  {title} {Vectorial polaritons in the quantum motion of a levitated
  nanosphere},\ }\href@noop {} {\bibfield  {journal} {\bibinfo  {journal}
  {Nature Physics}\ }\textbf {\bibinfo {volume} {17}},\ \bibinfo {pages} {1120}
  (\bibinfo {year} {2021})}\BibitemShut {NoStop}%
\bibitem [{\citenamefont {Romero-Isart}\ \emph {et~al.}(2010)\citenamefont
  {Romero-Isart}, \citenamefont {Juan}, \citenamefont {Quidant},\ and\
  \citenamefont {Cirac}}]{romero2010toward}%
  \BibitemOpen
  \bibfield  {author} {\bibinfo {author} {\bibfnamefont {O.}~\bibnamefont
  {Romero-Isart}}, \bibinfo {author} {\bibfnamefont {M.~L.}\ \bibnamefont
  {Juan}}, \bibinfo {author} {\bibfnamefont {R.}~\bibnamefont {Quidant}},\ and\
  \bibinfo {author} {\bibfnamefont {J.~I.}\ \bibnamefont {Cirac}},\ }\bibfield
  {title} {\bibinfo {title} {Toward quantum superposition of living
  organisms},\ }\href@noop {} {\bibfield  {journal} {\bibinfo  {journal} {New
  Journal of Physics}\ }\textbf {\bibinfo {volume} {12}},\ \bibinfo {pages}
  {033015} (\bibinfo {year} {2010})}\BibitemShut {NoStop}%
\bibitem [{\citenamefont {Dania}\ \emph {et~al.}(2022)\citenamefont {Dania},
  \citenamefont {Heidegger}, \citenamefont {Bykov}, \citenamefont {Cerchiari},
  \citenamefont {Araneda},\ and\ \citenamefont
  {Northup}}]{PhysRevLett.129.013601}%
  \BibitemOpen
  \bibfield  {author} {\bibinfo {author} {\bibfnamefont {L.}~\bibnamefont
  {Dania}}, \bibinfo {author} {\bibfnamefont {K.}~\bibnamefont {Heidegger}},
  \bibinfo {author} {\bibfnamefont {D.~S.}\ \bibnamefont {Bykov}}, \bibinfo
  {author} {\bibfnamefont {G.}~\bibnamefont {Cerchiari}}, \bibinfo {author}
  {\bibfnamefont {G.}~\bibnamefont {Araneda}},\ and\ \bibinfo {author}
  {\bibfnamefont {T.~E.}\ \bibnamefont {Northup}},\ }\bibfield  {title}
  {\bibinfo {title} {Position measurement of a levitated nanoparticle via
  interference with its mirror image},\ }\href
  {https://doi.org/10.1103/PhysRevLett.129.013601} {\bibfield  {journal}
  {\bibinfo  {journal} {Phys. Rev. Lett.}\ }\textbf {\bibinfo {volume} {129}},\
  \bibinfo {pages} {013601} (\bibinfo {year} {2022})}\BibitemShut {NoStop}%
\bibitem [{\citenamefont {Conangla}\ \emph {et~al.}(2019)\citenamefont
  {Conangla}, \citenamefont {Ricci}, \citenamefont {Cuairan}, \citenamefont
  {Schell}, \citenamefont {Meyer},\ and\ \citenamefont
  {Quidant}}]{PhysRevLett.122.223602}%
  \BibitemOpen
  \bibfield  {author} {\bibinfo {author} {\bibfnamefont {G.~P.}\ \bibnamefont
  {Conangla}}, \bibinfo {author} {\bibfnamefont {F.}~\bibnamefont {Ricci}},
  \bibinfo {author} {\bibfnamefont {M.~T.}\ \bibnamefont {Cuairan}}, \bibinfo
  {author} {\bibfnamefont {A.~W.}\ \bibnamefont {Schell}}, \bibinfo {author}
  {\bibfnamefont {N.}~\bibnamefont {Meyer}},\ and\ \bibinfo {author}
  {\bibfnamefont {R.}~\bibnamefont {Quidant}},\ }\bibfield  {title} {\bibinfo
  {title} {Optimal feedback cooling of a charged levitated nanoparticle with
  adaptive control},\ }\href {https://doi.org/10.1103/PhysRevLett.122.223602}
  {\bibfield  {journal} {\bibinfo  {journal} {Phys. Rev. Lett.}\ }\textbf
  {\bibinfo {volume} {122}},\ \bibinfo {pages} {223602} (\bibinfo {year}
  {2019})}\BibitemShut {NoStop}%
\bibitem [{\citenamefont {Wu}\ \emph {et~al.}(2014)\citenamefont {Wu},
  \citenamefont {Hryciw}, \citenamefont {Healey}, \citenamefont {Lake},
  \citenamefont {Jayakumar}, \citenamefont {Freeman}, \citenamefont {Davis},\
  and\ \citenamefont {Barclay}}]{PhysRevX.4.021052}%
  \BibitemOpen
  \bibfield  {author} {\bibinfo {author} {\bibfnamefont {M.}~\bibnamefont
  {Wu}}, \bibinfo {author} {\bibfnamefont {A.~C.}\ \bibnamefont {Hryciw}},
  \bibinfo {author} {\bibfnamefont {C.}~\bibnamefont {Healey}}, \bibinfo
  {author} {\bibfnamefont {D.~P.}\ \bibnamefont {Lake}}, \bibinfo {author}
  {\bibfnamefont {H.}~\bibnamefont {Jayakumar}}, \bibinfo {author}
  {\bibfnamefont {M.~R.}\ \bibnamefont {Freeman}}, \bibinfo {author}
  {\bibfnamefont {J.~P.}\ \bibnamefont {Davis}},\ and\ \bibinfo {author}
  {\bibfnamefont {P.~E.}\ \bibnamefont {Barclay}},\ }\bibfield  {title}
  {\bibinfo {title} {Dissipative and dispersive optomechanics in a nanocavity
  torque sensor},\ }\href {https://doi.org/10.1103/PhysRevX.4.021052}
  {\bibfield  {journal} {\bibinfo  {journal} {Phys. Rev. X}\ }\textbf {\bibinfo
  {volume} {4}},\ \bibinfo {pages} {021052} (\bibinfo {year}
  {2014})}\BibitemShut {NoStop}%
\bibitem [{\citenamefont {Purdy}\ \emph {et~al.}(2010)\citenamefont {Purdy},
  \citenamefont {Brooks}, \citenamefont {Botter}, \citenamefont {Brahms},
  \citenamefont {Ma},\ and\ \citenamefont
  {Stamper-Kurn}}]{PhysRevLett.105.133602}%
  \BibitemOpen
  \bibfield  {author} {\bibinfo {author} {\bibfnamefont {T.~P.}\ \bibnamefont
  {Purdy}}, \bibinfo {author} {\bibfnamefont {D.~W.~C.}\ \bibnamefont
  {Brooks}}, \bibinfo {author} {\bibfnamefont {T.}~\bibnamefont {Botter}},
  \bibinfo {author} {\bibfnamefont {N.}~\bibnamefont {Brahms}}, \bibinfo
  {author} {\bibfnamefont {Z.-Y.}\ \bibnamefont {Ma}},\ and\ \bibinfo {author}
  {\bibfnamefont {D.~M.}\ \bibnamefont {Stamper-Kurn}},\ }\bibfield  {title}
  {\bibinfo {title} {Tunable cavity optomechanics with ultracold atoms},\
  }\href {https://doi.org/10.1103/PhysRevLett.105.133602} {\bibfield  {journal}
  {\bibinfo  {journal} {Phys. Rev. Lett.}\ }\textbf {\bibinfo {volume} {105}},\
  \bibinfo {pages} {133602} (\bibinfo {year} {2010})}\BibitemShut {NoStop}%
\end{thebibliography}%

\end{document}